\documentclass[rmp,noshowpacs]{revtex4}
\usepackage{amsfonts,amssymb,latexsym,amsmath,amscd}
\usepackage{graphicx}
\usepackage{times,mathptm}
\usepackage[ps2pdf,colorlinks]{hyperref}
\usepackage{anysize}
\marginsize{1in}{1in}{1in}{1in}

\newcommand{\ket}[1]{|#1\rangle}
\newcommand{\bra}[1]{\langle#1|}
\newcommand{\M}[1]{\ket{\sf m#1}}

\newcommand{\mxA}{A}    
\newcommand{\mxB}{B}    
\newcommand{\mxH}{H}    
\newcommand{\mxI}{I}    
\newcommand{\mxQ}{Q}    
\newcommand{\mxR}{R}    
\newcommand{\mxS}{S}    
\newcommand{\mxT}{T}    
\newcommand{\mxU}{U}    
\newcommand{\mxV}{V}    
\newcommand{\mxVecs}{X} 
\newcommand{\mxW}{\omega}    
\newcommand{\conj}[1]{\overline{#1}}
\newcommand{\transp}[1]{{#1}^T}
\newcommand{\conjtransp}[1]{#1^\dagger}
\newcommand{\bea}{\begin{eqnarray*}}
\newcommand{\eea}{\end{eqnarray*}}
\newcommand{\bmx}{\left( \begin{array}}
\newcommand{\emx}{\end{array} \right)}

\newenvironment{example}
    {
    \smallskip
    \refstepcounter{theorem}
    \noindent
    {\bf Example \Roman{section}.\arabic{theorem}} \ \ }
    {
    \smallskip}

\newenvironment{remark}
    {
    \smallskip
    \refstepcounter{theorem}
    \noindent
    {\bf Remark \Roman{section}.\arabic{theorem}} \ \ }
    {\hspace*{\fill}
    \smallskip}

\newenvironment{definition}
    {
    \smallskip
    \refstepcounter{theorem}
    \noindent
    {\bf Definition \Roman{section}.\arabic{theorem}} \ \ }
    {\hspace*{\fill}{\ }
    \smallskip}

    {
    \smallskip
    \refstepcounter{theorem}
    \noindent
    {\bf Definition \Alph{section}.\arabic{theorem}} \ \ }
    {\hspace*{\fill}{\ }
    \smallskip}

    {
    \smallskip
    \refstepcounter{theorem}
    \noindent
    {\bf Scholium \Roman{section}.\arabic{theorem}} \it \ \ }
    {\hspace*{\fill}{\ }
    \smallskip}

\newenvironment{proof}[1][]
    {
    \noindent
    {\bf Proof{#1}:  }
    }
    {\hspace*{\fill}{$\Box$}\smallskip}

    {
    \noindent
    {\bf Sketch{#1}:  }
    }
    {\hspace*{\fill}{$\Box$}\smallskip}

    {
    \noindent
    {\bf Reference:  }
    }
    {\hspace*{\fill}{$\odot$}\smallskip}

\newtheorem{theorem}{Theorem}[section]
\newtheorem{proposition}[theorem]{Proposition}
\newtheorem{lemma}[theorem]{Lemma}
\newtheorem{corollary}[theorem]{Corollary}

\begin{document}

\title{Time Reversal and $n$-qubit Canonical Decompositions}

\author{Stephen S. Bullock}
\affiliation{Mathematical and Computational Sciences 
Division, \\
National Institute of Standards and Technology, Gaithersburg, MD,  20899}
\email{stephen.bullock@nist.gov}
\author{Gavin K. Brennen}
\affiliation{Atomic Physics Division, \\ 
National Institute of Standards and Technology, Gaithersburg, MD,  20899}
\email{gavin.brennen@nist.gov}
\author{Dianne P. O'Leary}
\affiliation{MCSD,
National Institute of Standards and Technology, \\
and Department of Computer Science, 
University of Maryland, College Park, MD,  20742}
\email{oleary@cs.umd.edu}

\begin{abstract}
On pure states of $n$ quantum bits, the concurrence entanglement monotone
returns the norm of the inner product of a pure state with its spin-flip.
The monotone vanishes for $n$ odd, but for $n$ even there is an explicit
formula for its value on mixed states, i.e. a closed-form
expression computes the minimum over all ensemble decompositions of a given
density.  For $n$ even a matrix decomposition $v=k_1 a k_2$
of the unitary group is explicitly computable and allows for study
of the monotone's dynamics.  The side factors $k_1$ and $k_2$ of
this Concurrence Canonical Decomposition (CCD) are concurrence symmetries,
so the dynamics reduce to consideration of the $a$ factor.  This
unitary $a$ is diagonal on a basis of GHZ-like states, with dynamics
of the entire state space depending on the relative phases within $a$.
In this work, we provide an explicit numerical algorithm computing
$v=k_1 a k_2$ for $n$ odd.  Further, in the odd case we lift the monotone
to a two-argument function.  The concurrence capacity of $v$ 
according to the double argument lift may be nontrivial for $n$ odd and
reduces to the usual concurrence capacity in the literature
for $n$ even.  The generalization may also be studied using the CCD, leading
again to maximal capacity for most unitaries.  The capacity of
$v \otimes I_2$ is at least that of $v$, so odd-qubit capacities have
implications for even-qubit entanglement.  The generalizations require
considering the spin-flip as a time reversal symmetry operator in Wigner's
axiomatization, and the original Lie algebra homomorphism defining the CCD may
be restated entirely in terms of this time reversal.  
The polar decomposition related to the CCD then writes any unitary evolution
as the product of a time-symmetric and time-antisymmetric evolution
w.r.t. the spin-flip.
\emph{En route}
we observe a Kramers' nondegeneracy:  the existence of a nondegenerate
eigenstate of any
\emph{time reversal symmetric} $n$-qubit Hamiltonian demands (i) $n$ even and
(ii) maximal concurrence of said eigenstate.  We provide examples of
how to apply this work to study the kinematics 
and dynamics of entanglement in spin chain Hamiltonians.
\end{abstract}

\maketitle

\noindent
\begin{center}
\hfill{\space}
{\bf AMS(MOS) subj:} 81P68, 
22E70, 
65F25 
\end{center}

\noindent
{\bf Running title:}  Time Reversal and Canonical Decompositions

\section{Introduction}

The entanglement theory of two quantum bits is now well understood.
For let $\rho$ be a mixed two-qubit quantum state, described  
by a $4 \times 4$ Hermitian density matrix.  Hill and Wootters
\cite{Hill:97} describe all classes of $\rho$ up to
evolution by unitaries in terms of the concurrence.  This concurrence is
explicitly a function of the eigenvalues of 
$\rho (\sigma^y)^{\otimes 2} \overline{\rho} (\sigma^y)^{\otimes 2}$,
where the factor $\tilde{\rho}=(\sigma^y)^{\otimes 2} 
\overline{\rho} (\sigma^y)^{\otimes 2}$ may be interpreted as the
spin-flip of $\rho$.  Further, for pure states
the entropy of the partial trace 
to either one-qubit subsystem is a one-to-one function of the concurrence,
so that both measures agree as to which two-qubit states are more or
less entangled.  Local states (tensors) are unentangled, while states locally
equivalent to Bell states have maximal entropy and concurrence.

For other systems, entanglement theory is more complicated.  Even for two
$d$-level systems (qudits,) it is not typical to use a single function to
quantify entanglement \cite{Vidal:00}, and research into
generalized concurrences continues \cite{Gour:04}.  Instead we focus on
the multi-partite qubit case.  The key point is that it is not
sensible in $n$-qubits to speak of a \emph{unique} maximally entangled state.
More precisely, suppose now $\rho$ is a $2^n \times 2^n$ Hermitian
density matrix describing a mixed $n$-qubit state.  A unitary
evolution is given by a $2^n \times 2^n$ unitary matrix, say
$v$, with the evolution being $\rho\mapsto v \rho v^\dagger$.  A partial
ordering of such $\rho$ as more or less entangled follows by stipulating
that (i) for $v= \otimes_{j=1}^n v_j$ local unitary, 
$\rho$ and $v \rho v^\dagger$ are equally entangled, 
while (ii) the $\rho$ becomes no more entangled on average after applying
any sequence of local measurements and local unitaries, i.e. after applying
local completely-positive maps.
More entangled is a partial order which has distinct
maximal elements for $n \geq 3$.  For example,
in three qubits, two states which are maximally entangled yet
locally inequivalent are given as follows \cite{DVC:01}:
\begin{equation}
\ket{\mbox{GHZ}} = (1/\sqrt{2}) \big[ \ket{000} + \ket{111} \big], \quad
\quad \quad \ket{\mbox{W}} = (1/\sqrt{3}) \big[ \ket{001}+\ket{010}+\ket{100}
\big] 
\end{equation}
There are nine distinct maxima of the partial order
in four qubits \cite{VDMD:02},
and strong theoretical evidence suggests that the number of such
\emph{entanglement types} grows quite rapidly with $n$
(e.g. \cite{Miyake:02}.)

To quantify multi-partite entanglement,
one often uses functions known as entanglement monotones 
\cite{Vidal:00,Barnum:01}.
All such monotones must vanish on any local state.
A monotone might also vanish on certain entangled states but 
definitively reports that
a state is not local should its value be nonzero. 
The value on a mixed state $\rho$ is defined to be the minimum over
all ensemble decompositions of $\rho$ of the ensemble weighted-average.
A monotone is convex on density matrices, since
entanglement does not increase under mixing of states.  
Monotones are also nonincreasing {\em on average} 
under local quantum operations and classical communication.
Among popular monotones are Meyer's $Q$-measure \cite{Meyer:02,Brennen:03},
the Schmidt measure \cite{Eisert:01},
and certain polynomial invariants \cite{Barnum:01} of eigenvalues
of density matrices representing stochastic mixtures of pure data states.

The $n$-qubit concurrence is an entanglement monotone.  
To define the monotone, we first note that \emph{throughout}
$\mho$ refers to the spin-flip of the $n$-qubit state space.
Concurrence for a pure
state \cite{Wong:01} is the component on a pure state of its spin-flip:
\begin{equation}
\label{eq:concurrence}
C_n (\ket{\psi})  =  \big| \langle \psi | \mho | \psi \rangle \big|/
\langle \psi | \psi \rangle, \mbox{ where }
\mho \ket{\psi} = \overline{ (-i \sigma^y)^{\otimes n} \ket{\psi}} =
(-i \sigma^y)^{\otimes n} \overline{\ket{\psi}}
\end{equation}
The concurrence of an $n$ qubit state with $n$ odd vanishes identically.
This monotone is noteworthy for two reasons.  First, there is an 
explicit, computable
closed-form expression for the minimum $C_n(\rho)$ which is again defined
in terms of the eigenvalues of 
$\rho \tilde{\rho}= \rho (\sigma^y)^{\otimes n} \overline{\rho} 
(\sigma^y)^{\otimes n}$ \cite{Uhlmann:00,BrennenBullock:04}.
Second, in the context of concurrence dynamics we may study entanglement
dynamics.  This paper concerns itself with the latter topic, and we henceforth
consider only pure states and unitary maps.

The primary mathematical tool used in this paper is the Concurrence 
Canonical Decomposition (CCD.)  This is discussed in 
detail in \S \ref{sec:background}.
Briefly, it is a way to decompose a unitary on $n$ qubits into a factor that 
changes concurrence and factors that do not. Let 
$v:\mathcal{H}_n \rightarrow \mathcal{H}_n$ be a unitary evolution.
Consider the CCD $v=k_1 a k_2$ \cite{BullockBrennen:03}.
Now $k_1$ and $k_2$ are symmetries of the concurrence, 
reducing concurrence dynamics to the
second factor.  This $a$ factor applies relative phases to a basis
of GHZ-like states.
Such phases are not unique due to choices of diagonalization
while computing the CCD, but the spectrum $\mbox{spec}(a^2)$
is uniquely determined by $v$.  Moreover, the two-qubit test for
maximal entanglement capacity
\cite{Zhang:03} generalizes to $n$ qubit concurrence
capacities if $n$ is even:
\begin{quotation}
Let $v=k_1 a k_2$ be a CCD of $v$.
Consider $\mbox{spec}(a^2)$ as a subset of the unit circle.  Then for
$n=2p$, there is a $\ket{\psi}\in \mathcal{H}_n$ with
$C_n(\ket{\psi})=0$ and $C_n (v\ket{\psi})=1$ if and only if
$0$ is within the complex hull of $\mbox{spec}(a^2)$.
\cite{Zhang:03,BullockBrennen:03}
\end{quotation}
Also, for even $n$ there is an explicit numerical
algorithm for computing the CCD
and hence $\mbox{spec}(a^2)$ \cite{BullockBrennen:03}.

This work presents three new results.  The first is an extension of
concurrence capacities to the case $n$ odd.  For $n$ even, the concurrence
symmetry group $K$ to which $k_1$, $k_2$ belong is up to a similarity
transform an orthogonal group.  For $n$ odd, $K$ is not orthogonal
but symplectic, $a$ has repeat eigenvalues, and
$C_{2p-1}(\ket{\psi})=0$ for all $\ket{\psi}$.  Nonetheless, we define
a two-argument lift of the usual concurrence, say 
$\mathcal{C}(\ket{\phi},\ket{\psi})$.  (See Equation \ref{eq:conc_form}.)
Suppose we
define the amount of concurrence an odd-qubit unitary $v$ creates to be
\begin{equation}
\kappa(v) \ = \ \mbox{max }\big\{ \; 
\mathcal{C}(v\ket{\phi},v\ket{\psi}) \; ; \;
\mathcal{C}(\ket{\phi},\ket{\psi}) = 0 \; \big\}
\end{equation}
This generalized capacity has the following properties:
\begin{itemize}
\item  For $n$ even, the one-argument
concurrence capacity and the two-argument capacity of $v$ coincide.
\item  For $n$ odd, often $\kappa_n(v) \neq 0$ for the pairwise capacity
despite $C_{n}(\ket{\psi}) \equiv 0$.  Further, $\kappa_n(v)=1$
if and only if $0$ lies within the convex hull of $\mbox{spec}(a^2)$
for any CCD by $v=k_1 a k_2$.
\item  \emph{Concurrence capacity monotonicity:}  Using double argument
capacities, the capacity of $v \otimes I_2$ is always at least that of $v$.
\end{itemize}
Hence there exists a theory of odd-qubit concurrence dynamics, even
though concurrence vanishes identically (on the diagonal) in odd qubits.

Second, we present an explicit numerical algorithm for computing the
odd-qubit CCD.  Various matrix logarithms must be computed, after which
one invokes work in the numerical analysis literature \cite{DongarraEtAl:84}
to diagonalize a time reversal symmetric Hamiltonian using
symplectic matrices.

We close with the third observation, which we will
refer to as \emph{Kramers' nondegeneracy:}
\begin{quotation}
On the $n$-quantum bit state space, suppose that a $\mho$-time reversal
symmetric Hamiltonian $H$ has a nondegenerate eigenstate $\ket{\lambda}$.
Then (i) $n$ is even and (ii) $C_n(\ket{\lambda})=1$.  In particular,
$\ket{\lambda}$ is entangled, i.e. $\ket{\lambda} \not\in
(\mathcal{H}_1)^{\otimes n}$.
\end{quotation}
The proof follows from viewing $\mho$ as a time reversal symmetry
operator in Wigner's axiomatization, a point of view which also
simplifies the derivation of the CCD.
Kramers' nondegeneracy leads one to wonder whether 
useful entangled states may be
produced by cooling the system of a $\mho$-time reversal symmetric
Hamiltonian.  We consider the perturbative stability of this entanglement
while breaking the time reversal symmetry here, while the thermal
stability of the Kramers' nondegeneracy for the Ising model is
considered elsewhere \cite{BrennenBullock:04}.

\section{Background and Prior Work}

\label{sec:background}

Since our key tool is a generalized Canonical Decomposition
\cite{BullockBrennen:03}, we review the Canonical Decomposition
literature.  The two-qubit Canonical Decomposition (CD) states that 
any two-quantum bit unitary evolution $v$,
i.e. any $4 \times 4$ unitary matrix $v$, may be written:
\begin{equation}
\label{eq:cd}
v = \mbox{e}^{i \varphi} (u_1 \otimes u_2) a (u_3 \otimes u_4)
\end{equation}
Here $u_1,u_2,u_3,u_4$ are one-qubit $(2\times 2)$ unitary matrices,
which may be chosen to have determinant one.  The unitary $a$ is diagonal
in the Bell basis and may be thought of as applying relative phases to
this basis.  However, it is better computationally to think of $a$
as phasing the \emph{magic basis}  \cite{Bennett:96,LKea:01} instead:
\begin{equation}
\begin{array}{lclclcl}
\M{0} & = & (\ket{00} + \ket{11})/\sqrt{2}, & \quad \quad &
\M{1} & = & (\ket{01} - \ket{10})/\sqrt{2} \\
\M{2} & = & (i\ket{00} - i\ket{11})/\sqrt{2}, & \quad \quad &
\M{3} & = & (i\ket{01} + i\ket{10})/\sqrt{2} \\
\end{array}
\end{equation}
Let $E$ be defined by $E\ket{j}=\M{j}$,
and let $SU(2^n)$ denote the Lie group of determinant
one $2^n \times 2^n$ unitary matrices, $SO(2^n)$ denotes
determinant one orthogonal matrices, and 
$D(2^n)$ denotes the diagonal $2^n \times 2^n$ unitary
matrices.  A diagonalization argument
shows $SU(4)=SO(4) \; D(4) \; SO(4)$.  Moreover, the magic basis has
the property that $E^\dagger SU(2) \otimes SU(2) E = SO(4)$, i.e.
\emph{determinant one} tensors have real matrix coefficients in the basis.
Thus the canonical decomposition may be computed
by transforming the diagonalization through $E$:
\begin{equation}
SU(4) \ = \ [E SO(4) E^\dagger][E D(4) E^\dagger][E SO(4) E^\dagger] \ = \ 
SU(2) \otimes SU(2) \; (E D(4) E^\dagger) \; SU(2) \otimes SU(2)
\end{equation}
We next provide a brief account and references for the best known
applications and generalizations of the CD.

Makhlin \cite{Makhlin:00} anticipates the Canonical Decomposition by
directly computing that the double cosets 
$[SU(2)\otimes SU(2)] \backslash SU(4)/ [SU(2)\otimes SU(2)]$
are parametrized by three real parameters, the number of
parameters in $a$ given $\mbox{det}(a)=1$.
The CD appears explicitly in Kraus and Cirac 
\cite{KrausCirac:01}.  In an important paper,
Khaneja, Brockett, and Glaser point out that one may view the
CD as an example of the $G=KAK$ decomposition theorem for
$G=SU(4)$, $K=SU(2) \otimes SU(2)$, and $A=\Delta$ the commutative Lie group
that phases the magic (or Bell) basis \cite{KBG:01}.
They also consider
the matrix factorization from the point of view of control theory
in order to compute minimum times for applying a given two-qubit unitary
evolution.  Zhang, Vala, Sastry, and Whaley 
made use of this observation to
describe which $4\times 4$ unitaries $v$ are equivalent up to 
tensors of one-qubit rotations.  The factor $a \in \Delta$ is not unique
but depends on choices of diagonalization, and these are described
geometrically using Weyl chambers.
Specifically, the Weyl group orbit
of any $a$ produces all possible $a$, 
and each orbit intersects the Weyl chamber once.
For $G=SU(4)$, the Weyl chamber is a tetrahedron \cite{Zhang:03}.  Finally, 
the terms Canonical
Decomposition and magic basis are by now standard, and there are
published surveys (e.g. \cite[\S II.B]{ChildsEtAl:03}.)  
The timing arguments of Khaneja et. al.
\cite{KBG:01} have also recently been experimentally verified
in liquid-state NMR \cite{Juha_exp:04}.

There are many applications of the two-qubit CD.  In addition
to timing as above, they include
(i) studying the entanglement capacity of two-qubit operations \cite{Zhang:03},
(ii) building efficient (small) quantum circuits in two qubits
\cite{Bullock:03,VatanWilliams:03,Vidal:03,Shende:03}, and (iii)
classifying which two-qubit computations require fewer than
average multiqubit interactions \cite{Vidal:03,Shende:03}.

Besides the CCD \cite{BullockBrennen:03}, 
there is another $n$-qubit generalization of the
Canonical Decomposition due to Khaneja and Glaser \cite{KG:01}.
It is also defined in terms of a $G=KAK$ decomposition.
Label $N=2^n$ for the remainder.  The type of a $G=KAK$
decomposition follows from a classification theorem of
Cartan involutions and determines the groups $K$ and $A$ up
to Lie isomorphism.  (The classification appears in Helgason
\cite[pg.518]{Helgason:01}.  See ibid. for details.)  
Given $G=SU(N)$, the three possible types demand
$K \cong SO(N)$ (type {\bf AI},) 
$K \cong Sp(N/2)$ a symplectic group (type {\bf AII},)
or $K \cong S[U(p) \oplus U(q)]$ for $p+q=N$ a block unitary (type {\bf AIII}).
In the {\bf AII} case, the structure of the $A$ group also demands
any $a \in A$ has even-degenerate eigenvalues.
The two-qubit Canonical Decomposition
is type {\bf AI}, and indeed the similarity transform by $E$
shows $SU(2) \otimes SU(2) \cong SO(4)$.
The CCD alternates {\bf AI} and {\bf AII}
as $n$ is even or odd.  The KGD of Khaneja and Glaser technically
contains two $G=KAK$ decompositions, the first of which is type
{\bf AIII} for $n>2$.  In fact, the KGD is similar 
to the Cosine Sine Decomposition (CSD) of numerical 
linear algebra \cite{Bullock:04} and so may be computed numerically.
Physically, the $K\cong S[U(N/2)\oplus U(N/2)]$ group of the KGD
may be viewed as those unitaries commuting with measurements 
in the $z$-basis of the least significant qubit, 
i.e. commuting with $I_{N/2} \otimes \sigma^z$.

We next recall notation from quantum computing.
The one-qubit state space is
$\mathcal{H}_1 = \mathbb{C} \{ \ket{0}\} \oplus \mathbb{C} \{ \ket{1}\}$.
For $n$ quantum bits,
\(\mathcal{H}_n = (\mathcal{H}_1)^{\otimes n} = \mathcal{H}_1 \otimes \cdots
\otimes \mathcal{H}_1\).  (See \cite{NielsenChuang:00}.)
A local state $\ket{\psi}$ is any state which may be written as
$\otimes_{j=1}^n \ket{\psi_j}$ for $\ket{\psi_j} \in \mathcal{H}_1$, while
an \emph{entangled} state is any state which is not local.  
Notations such as e.g. $\ket{7}$ refer not to the state
of a qudit but rather to a multiqubit state, e.g. 
$\ket{111}=\ket{1}\otimes\ket{1}\otimes\ket{1}$.
The $n$-concurrence of Equation \ref{eq:concurrence}
is an entanglement monotone \cite{BullockBrennen:03}. 
Besides the well-known two-qubit concurrence
\cite{Hill:97}, even qubit concurrences
$n$-qubits \cite{Wong:01} \cite[Eq.62]{Scott:03} \cite{BullockBrennen:03}
have also been studied.
Since the single-argument concurrence vanishes for $n$ odd,
we introduce a two-argument generalization.  

For $\mho$ per Equation \ref{eq:concurrence}, the \emph{concurrence
bilinear form} \cite{BullockBrennen:03} is the map 
$\mathcal{C}_n:\mathcal{H}_n \times \mathcal{H}_n \rightarrow \mathbb{C}$ 
given by 
\begin{equation}
\label{eq:conc_form}
\mathcal{C}_n(\ket{\phi},\ket{\psi})= \overline{\bra{\phi}\mho\ket{\psi}}
\end{equation}
The complex conjugate forces the two-argument function to be complex
bilinear rather than complex bi-antilinear,
and the concurrence monotone is the norm of the
form on the diagonal:
$C_n(\ket{\phi})=|\mathcal{C}_n(\ket{\phi},\ket{\phi})|$.  The bilinear form 
$\mathcal{C}_n$
is symmetric for $n$ even and antisymmetric for $n$ odd, which causes
vanishing of the monotone \emph{but not the form} in the odd-qubit case.

The CD is an example of the $G=KAK$
decomposition theorem 
\cite[thm8.6,\S VII.8]{Helgason:01} for $G=SU(N)$.
This theorem produces a decomposition of a reductive
Lie group $G$ for any $\theta$, $\mathfrak{a}$ as follows:
\begin{itemize}
\item The map
\hbox{$\theta:\mathfrak{g}\rightarrow \mathfrak{g}$} for
$\mathfrak{g}=\mbox{Lie}(G)$ is a Cartan involution
\cite[\S X.6.3,pg.518]{Helgason:01}.  
By definition\footnote{Some authors only use the term
Cartan involution in the case that $\mathfrak{g}$ is a noncompact
Lie algebra.  In their terminology, this definition of
a Cartan involution on the Lie algebra of a compact group, 
e.g. $\mathfrak{su}(N)$,
is the image of a
Cartan involution of a noncompact Lie algebra through symmetric duality
$(\mathfrak{g}=\mathfrak{k}\oplus \mathfrak{p}) \longleftrightarrow
(\mathfrak{g}^{\mbox{dual}}=\mathfrak{k} \oplus i \mathfrak{p})$.},
(i) $\theta^2 = {\bf 1}_{\mathfrak{g}}$ and (ii)
$\theta[X,Y]=[\theta X, \theta Y]$ for all $X,Y \in \mathfrak{g}$.
As is standard, we write 
$\mathfrak{g}=\mathfrak{p} \oplus \mathfrak{k}$ for the
decomposition of $\mathfrak{g}$ into the $-1$ and $+1$ eigenspace
of $\theta$.  
\item Given $\theta$,
$\mathfrak{a} \subset \mathfrak{p}$ is a 
commutative subalgebra which
is maximal commutative in $\mathfrak{p}$.
\end{itemize}
Note that $\mathfrak{k}$ is
closed under the Lie bracket, while this is trivially true
for $\mathfrak{a}$.  Thus the exponential of each is a group.
Label $K = \mbox{exp } \mathfrak{k}$, $A = \mbox{exp } \mathfrak{a}$,
where for linear $G \subset GL(n,\mathbb{C})$ the exponential may be
interpreted as a matrix exponential.  The theorem then asserts that
$G = KAK = \{ k_1 a k_2 \; ; \; k_1,k_2 \in K, a \in A \}$.

The CD is seen to be an example as follows, cf. \cite{KBG:01}.  
Take
\hbox{$\theta: \mathfrak{su}(4) \rightarrow \mathfrak{su}(4)$}
by $\theta(X)=(-i \sigma^y)^{\otimes 2} \overline{X}
(-i \sigma^y)^{\otimes 2}$ and
\hbox{\( \mathfrak{a} = \mbox{span}_{\mathbb{R}} 
\big\{ 
\ \ i \ket{0}\bra{0}-i\ket{1}\bra{1}-i\ket{2}\bra{2}+i\ket{3}\bra{3}, \ 
i\ket{0}\bra{3}+i\ket{3}\bra{0},\ i\ket{1}\bra{2}+i\ket{2}\bra{1} 
\ \ \big\} \)}.
Extending these choices to $n$ qubits produces the CCD:

\vbox{
\begin{definition}[CCD,\cite{BullockBrennen:03}]
\label{def:CCD}
Define $\theta:\mathfrak{su}(N) \rightarrow \mathfrak{su}(N)$
by $\theta(X)=[(-i\sigma^y)^{\otimes n}]^\dagger 
\overline{X} (-i\sigma^y)^{\otimes n}$.  
Then $\mathfrak{k}$ denotes the $+1$-eigenspace of $\theta$ while
$\mathfrak{p}$ denotes the $-1$-eigenspace.  Finally, in case
$n$ is even we define
\begin{equation}
\begin{array}{lcl}
\mathfrak{a} & = &
\mbox{span}_{\mathbb{R}}
\big( \{ \ 
i \ket{j}\bra{j}+i \ket{N-j-1}\bra{N-j-1} -i
\ket{j+1}\bra{j+1}-i\ket{N-j-2}\bra{N-j-2} \ ; \ 
0 \leq j \leq 2^{n-1}-2 \ 
\} \\
& & \quad \quad \quad \quad \sqcup \quad
\{ \ i \ket{j}\bra{N-j-1} + i\ket{N-j-1} \bra{j}  \ ; \
0 \leq j \leq 2^{n-1}-1 \ \} , \big) 
\end{array}
\end{equation}
with $A=\mbox{exp}\; \mathfrak{a}$.  In case $n$ odd, we drop the
second set:
\begin{equation}
\begin{array}{lcl}
\mathfrak{a} & = &
\mbox{span}_{\mathbb{R}}
\big( \{ \ 
i \ket{j}\bra{j}+i \ket{N-j-1}\bra{N-j-1} -i
\ket{j+1}\bra{j+1}-i\ket{N-j-2}\bra{N-j-2} \ ; \ 
0 \leq j \leq 2^{n-1}-2 \ 
\} \big) \\
\end{array}
\end{equation}
The \emph{Concurrence
Canonical Decomposition (CCD)} in $n$-qubits is the resulting
matrix decomposition $SU(N)=KAK$.  Note that $n$ may be even or odd.
\end{definition}
}

In an earlier work \cite{BullockBrennen:03}, 
computations in Dirac (bra-ket) notation show that
$\theta(X)$ is a Cartan involution and
$\mathfrak{a}$ is maximal-commutative in $\mathfrak{p}$.  
The $G=KAK$ theorem \cite[thm8.6,\S VII.8]{Helgason:01}
then shows that the CCD exists.  
Further, the CCD may be computed numerically
in the even qubit case \cite{BullockBrennen:03}.  

The CCD is a useful tool for studying concurrence
capacities since $K=\mbox{exp}(\mathfrak{k})$ consists
of symmetries of the concurrence form of Equation \ref{eq:conc_form}, 
where $\mathfrak{k}$ is
given per Definition \ref{def:CCD} \cite{BullockBrennen:03}.
\begin{equation}
\label{eq:K_is_symmetry_group}
(v \in K) \Longleftrightarrow [ \; \mathcal{C}_n(v\ket{\phi},v\ket{\psi})=
\mathcal{C}_n(\ket{\phi},\ket{\psi}) \mbox{ for all } \ket{\phi},\ket{\psi}
\in \mathcal{H}_n \; ]
\end{equation} 
In particular, the above may be used to verify
that $SU(2)^{\otimes n} \subseteq K$ as a subgroup of large codimension.
One explanation for the fact that 
$K$ alternates between orthogonal and symplectic
groups is to note that the form $\mathcal{C}_n$ is symmetric or
antisymmetric as $n$ is even or odd \cite{BullockBrennen:03}.  Another
outlook, illustrated in \S \ref{sec:time}, is that the spin-flip
$\mho$ is a bosonic or fermionic time reversal symmetry operator
as $n$ is even or odd, i.e. $\mho^{-1}=(-1)^n \mho$.

\section{Odd-qubit concurrence capacities}

The main results of this section are summarized in
Theorem \ref{thm:kappas}.  Each is proven in turn.

\subsection{Double-argument capacities generalize single-argument capacities}
\label{subsec:double_argument}

To begin, we introduce a pairwise concurrence
capacity $\kappa_n(v)$ and denote earlier concurrence capacities
\cite{BullockBrennen:03} with a tilde.
\begin{equation}
\left\{
\begin{array}{lcl}
\tilde{\kappa}_n(v) & = & \mbox{max} \{ \; C_n(v \ket{\psi}) \; ; \;
\bra{\psi} \psi \rangle =1, C_n(\ket{\psi})=0 \; \} \\
\kappa_n(v) & = & \mbox{max} \{ \; |\mathcal{C}_n(v \ket{\phi}, v\ket{\psi})| 
\; ; \;
\bra{\phi} \phi \rangle=\bra{\psi} \psi \rangle = 1,
\mathcal{C}_n(\ket{\phi},\ket{\psi})=0 \; \} \\
\end{array}
\right.
\end{equation}
Due to Equation \ref{eq:K_is_symmetry_group},
any CCD of a unitary $v=k_1 a k_2$ implies
$\tilde{\kappa}_n(v)=\tilde{\kappa}_n(a)$ \cite{BullockBrennen:03} and
$\kappa_n(v)=\kappa_n(a)$.

\begin{proposition}
\label{prop:twokaps}
Suppose $n=2p$ is an even number of qubits.  Then 
$\kappa_n(v)=\tilde{\kappa}_n(v)$.
\end{proposition}

The proof requires certain results from the literature  
\cite{BullockBrennen:03,Zhang:03}.
\begin{itemize}
\item  There is an $n=2p$ qubit entangler $E_0$ so that for any
$k \in K$, $E_0 k E_0^\dagger$ is a real unitary matrix, i.e. orthogonal.
The columns of $E_0$ resemble $\ket{\mbox{GHZ}}$ states.
\item  For this $E_0$, any CCD $v=k_1 a k_2$ moreover
has $d= E^\dagger a E$ for $d=\sum_{j=0}^{N-1} d_j \ket{j} \bra{j}$ diagonal.
As $d$ is unitary diagonal, each $d_j$ is on the unit circle within
$\mathbb{C}$.
\item  The concurrence spectrum becomes $\lambda_c(v)=
\{ d_j^2 \}_{j=0}^{N-1}$.  Then
$\tilde{\kappa}_{2n}(v)=1$ if and only if
$0 \in \mathbb{C}$ lies within the convex hull of $\lambda_c(v)$,
a subset of the unit circle \cite[Lem.III.2]{BullockBrennen:03}.
\item  A corollary \cite[Scho.2.18]{BullockBrennen:03} of the symmetry
group theorem shows that $E_0$ also translates between $\mathcal{C}_n(-,-)$
and a simpler bilinear form:  
$\mathcal{C}_n(E_0z_1,E_0 z_2)=z_1^T z_2$.
\end{itemize}

\begin{example}
We use the CD to compute a two-qubit concurrence capacity.
Consider a family of
controlled-phase gates, e.g. $v(t)=\mbox{e}^{-it}\ket{0}\bra{0}+
\mbox{e}^{-it}\ket{1}\bra{1}+\mbox{e}^{-it}\ket{2}\bra{2}+
\mbox{e}^{3it}\ket{3}\bra{3}$ with $\mbox{det}[v(t)]=1$.  A possible CD is:
\begin{equation}
v(t)\ =\  (\mbox{e}^{-it \sigma^z} \otimes I_2) \;
\mbox{e}^{{it \sigma^z \otimes \sigma^z}} 
\; (I_2 \otimes \mbox{e}^{-i t \sigma^z})
\end{equation}
The central factor is a valid choice for $a$ in $v(t)=k_1 a k_2$,
since $\mbox{e}^{it \sigma^z \otimes \sigma^z}$ is
also diagonal in the magic basis.
Thus $\lambda_c[v(t)]=\mbox{spec}(\mbox{e}^{2it\sigma^z \otimes \sigma^z})=
\{ \mbox{e}^{2it},\mbox{e}^{2it},\mbox{e}^{-2it},\mbox{e}^{-2it} \}$.
Only for $t \in \frac{\pi}{4}\mathbb{Z}$ 
do we have $0$ within the convex
hull of $\lambda_c[v(t)]$, and the convex hull theorem asserts
$\tilde{\kappa}_2[v(\pi/4)]=1$.
Indeed, up to phase
$v(\pi/4)=\ket{0}\bra{0}+\ket{1}\bra{1}+\ket{2}\bra{2}-\ket{3}\bra{3}$.
Moreover, if $H=\frac{1}{\sqrt{2}}\left( \begin{array}{rr} 1 & 1 \\
1 & -1 \\ \end{array}\right)$ is the Hadamard gate \cite{NielsenChuang:00},
a standard identity converts $v(\pi/4)$ into the quantum controlled-not:
\begin{equation}
{\tt CNOT} = 
\ket{00}\bra{00}+\ket{01}\bra{01}+\ket{10}\bra{11}+\ket{11}\bra{10}
=(I_2 \otimes H) v(\pi/4) (I_2 \otimes H)
\end{equation}
Thus $v(\pi/4)$ carries an unentangled state to a maximally entangled
state, since ${\tt CNOT} (H \otimes I_2)\ket{00}=
{\tt CNOT} \frac{1}{\sqrt{2}} (\ket{00}+\ket{10})=
\frac{1}{\sqrt{2}}(\ket{00}+\ket{11})$.  More intricate examples in two-qubits 
\cite{Zhang:03,Shende:03} and an
even number of qubits \cite{SPIE} are available in the literature.
\end{example}

\vbox{
\begin{lemma}
\label{lem:pseudodot}
Suppose the number of qubits is even.  Let 
$z_1=\sum_{j=0}^{N-1} a_j \ket{j}$,
$z_2=\sum_{j=0}^{N-1} b_j \ket{j}$, and $z_3=\sum_{j=0}^{N-1} c_j \ket{j}$
throughout, and let $\lambda_c(v)=\{\lambda_j\}_{j=0}^{N-1}$.  
Then we have the following.
\begin{equation}
\label{eq:max_kappa}
\begin{array}{lcl}
\tilde{\kappa}_n(v)& = & \mbox{max } \{ |\sum_{j=0}^{N-1} c_j^2 \lambda_j|
\; ; \; z_3^\dagger z_3 = 1, z_3^T z_3 =0 \} \\
\kappa_n(v) & = & \mbox{max } \{ |\sum_{j=0}^{N-1} a_j b_j \lambda_j| \; ; \;
z_1^\dagger z_1 =  z_2^\dagger z_2 = 1, z_1^T z_2 = 0\} \\
\end{array}
\end{equation}
\end{lemma}
}

\begin{proof}[ \ of Lemma \ref{lem:pseudodot}]
The first equation appears in \cite{BullockBrennen:03};
cf. \cite{Zhang:03}.  For the second, take vectors 
$z_1, z_2$ and label $x=E_0 z_1$, $y=E_0 z_2$.  Then
\begin{equation}
[\mathcal{C}_n(x,y)=0] \Longleftrightarrow
[\mathcal{C}_n(E_0z_1,E_0z_2)=0] \Longleftrightarrow
[z_1^T z_2 = 0]
\end{equation}
Moreover, without loss of generality by choice of $z_1$, $z_2$, and
symmetry we may suppose $v=E_0 d E_0^\dagger$ for 
$d^2=\sum_{j=0}^{N-1} \lambda_j \ket{j}\bra{j}$.  Then
\(
\mathcal{C}_n(E_0dE_0^\dagger x, E_0dE_0^\dagger y) = 
\mathcal{C}_n(E_0dz_1, E_0dz_2) =
(z_1^T d^T) d z_2 = \sum_{j=0}^{N-1} a_j b_j \lambda_j
\).
\end{proof}

\begin{proof}[\ of Proposition \ref{prop:twokaps}]
Let $a_j$, $b_j$ be chosen
so as to maximize the expression for $\kappa_n(v)$
per Lemma \ref{lem:pseudodot}, i.e. 
$\kappa_n(v)=|\sum_{j=0}^{N-1} a_j b_j \lambda_j|$.  Now choose complex
numbers $c_j$ so that $c_j^2 = a_j b_j$, and put 
$z_3=\sum_{j=0}^{N-1} c_j \ket{j}$.  We note that $z_3^T z_3 = 0$.
Moreover, $z_3^\dagger z_3 \leq 1$, for
\begin{equation}
\sum_{j=0}^{N-1} |c_j|^2 = \sum_{j=0}^{N-1} |c_j^2| =
\sum_{j=0}^{N-1} |a_j b_j| \ \leq 
\ \sum_{j=0}^{N-1} \frac{1}{2} |a_j|^2 + \frac{1}{2} |b_j|^2  = 1
\end{equation}
Label $t^2=z_3^\dagger z_3$, noting $t^2 \leq 1$.  Then 
$(t^{-1} z_3)^\dagger (t^{-1} z_3)=1$, so by definition of
$\tilde{\kappa}_{2p}(v)$ we have
\begin{equation}
\kappa_n(v) \geq \tilde{\kappa}_n(v) \ \geq \ 
|\sum_{j=0}^{N-1} t^{-2} c_j^2 \lambda_j| \ = \ 
t^{-2} |\sum_{j=0}^{N-1} a_j b_j \lambda_j| \ = \ t^{-2} \kappa_n(v)
\end{equation}
Thus $t=1$ and hence $\kappa_n(v)=\tilde{\kappa}_n(v)$.
\end{proof}

\subsection{Monotonicity}

We next demonstrate concurrence capacity monotonicity,
i.e. that $j \mapsto \kappa_{n+j} (v \otimes I_2^{\otimes j})$ is
monotonic.

\begin{proposition}
\label{prop:monotonicity}
Let $n$ be either even or odd, $v \in SU(N)$ an $n$-qubit computation,
and let $I_2$ denote the trivial one-qubit computation.  Then
\( \kappa_{n+1}(v \otimes I_2) \ \geq \ \kappa_n(v) \).
\end{proposition}

\begin{proof}
Choose $\ket{\phi}$, $\ket{\psi}$ such that
$\kappa_n(v)= \mathcal{C}_n(v \ket{\phi}, v \ket{\psi})$ while
$\mathcal{C}_n(\ket{\phi},\ket{\psi})=0$.
Then $\ket{\phi} \otimes \ket{0}$ and $\ket{\psi} \otimes \ket{1}$
are a null-concurrent pair of $(n+1)$-qubit states:
\begin{equation}
\mathcal{C}_{n+1}(\ket{\phi} \otimes \ket{0}, \ket{\psi} \otimes \ket{1})\ = \
\overline{(\bra{\phi} \otimes \bra{0})} (-i \sigma^y)^{\otimes n+1} 
(\ket{\psi} \otimes \ket{1}) \ = \ 
[ \mathcal{C}_n (\ket{\phi},\ket{\psi}) ] (\overline{\bra{0}} 
(-i \sigma^y) \ket{1})
\end{equation}
Now $\overline{\bra{0}} (-i \sigma^y) \ket{1})=1$, so the above expression
is $[0](1)=0$.  A similar argument demonstrates that
\begin{equation}
\mathcal{C}_{n+1}[ (v \otimes I_2) (\ket{\phi}\otimes \ket{0}),
(v \otimes I_2) (\ket{\psi}\otimes \ket{1})] \ = \ 
[\mathcal{C}_n(v \ket{\phi}, v\ket{\psi})] \ [\mathcal{C}_1(\ket{0},\ket{1})]
\end{equation}
The second term of the product is one, while the first is
$\kappa_n(v)$.  Since we have exibited a pair for which
$v \otimes I_2$ raises the pairwise concurrence by at least
$\kappa_n(v)$.  Since $\kappa_{n+1}(v \otimes I_2)$
is the maximum over all null-concurrent pairs, while
$\ket{\phi}\otimes \ket{0}$, $\ket{\psi} \otimes \ket{1}$ is such,
we see $\kappa_{n+1}(v \otimes I_2) \geq \kappa_n(v)$.
\end{proof}

\subsection{Parity-independent concurrence spectra}
\label{subsec:spectra_&_AII}

We extend the maximal concurrence capacity condition of
Zhang et. al. and Bullock, Brennen \cite{BullockBrennen:03}
to odd-qubit systems.  The first step is a definition valid in either parity.

\begin{definition}
\label{def:conspec}
Let $v \in SU(N)$, $N=2^n$.
For $n$ of either parity, the concurrence spectrum $\lambda_c(v)$ is the set
$\lambda_c(v)= 
\mbox{spec}
\bigg( [(-i \sigma^y)^{\otimes n}]^\dagger v (-i \sigma^y)^{\otimes n} v^T
\bigg)$.  Viewing $v$ as an $\mathbb{R}$-linear map,
equivalently $\lambda_c(v)=\mbox{spec} \big( v \mho v^\dagger \mho^{-1})$.
\end{definition}

We briefly show this coincides with the definition of the
even-qubit concurrence spectrum  of the literature \cite{BullockBrennen:03}.  
The definition ibid. states that the concurrence spectrum is the
spectrum of $(E_0^\dagger v E_0) (E_0^\dagger v E_0)^T$.
Indeed, given $E_0 E^T_0 = (-i \sigma^y)^{\otimes n}$ per the classification
of $E$ with $E SO(N) E^\dagger=K$ ibid.,
\begin{equation}
\mbox{spec}\; (E_0^\dagger v E_0) (E_0^\dagger v E_0)^T \ = \ 
\mbox{spec}(E_0^\dagger v E_0 E_0^T v^T \overline{E_0}) \ = \ 
\mbox{spec}[(E_0 E_0^T)^\dagger v E_0 E_0^T v^T ] \ = \ 
\mbox{spec}[( -i \sigma^y)^{\otimes n} v (-i \sigma^y)^{\otimes n}v^T]
\end{equation}
In fact, the same argument shows that
$\lambda_c(v)$ is the spectrum $(E^\dagger v E) (E^\dagger v E)^T$
for any $E$ as above.  Cf. \cite{Makhlin:00}.

The odd-qubit case requires different similarity matrices,
say $F$ \cite{BullockBrennen:03}, which translate $K$ not into
an orthogonal group but rather a symplectic group per 
Equation \ref{eq:symplectic}.  For the
concurrence form $\mathcal{C}_n(-,-)$ for $n$-odd is antisymmetric, 
and symplectic rather than orthogonal groups are the appropriate symmetries
of antisymmetric bilinear forms (i.e. two-forms.)  For a standard similarity
matrix, we take

\begin{equation}
\begin{array}{l}
F_0 = \sum_{j=0}^{N/2-1} 
\ket{j}\bra{j}  +  \ket{N-j-1}\bra{j}
\; + \; \iota_j ( \ket{j}\bra{N/2+j} - \ket{N-j-1}\bra{N/2+j}), \mbox{ where }
\\
\{ \iota_j\}_{j=0}^{N/2-1} \subset \{\pm 1\} \mbox{   by   }
(-i\sigma^y)^{\otimes n} = \sum_{j=0}^{N/2-1}
\iota_j (\ket{N-j-1}\bra{j}-\ket{j}\bra{N-j-1}) \\
\end{array}
\end{equation}
Also, label throughout $J_N=(-i\sigma^y)\otimes I_{N/2}$.  
Before showing that $F_0$ translates $K$ into the standard symplectic
group, we show that $F_0$ carries $\mathcal{C}(-,-)$ to the standard
two-form $\mathcal{A}(-,-)$.

\begin{lemma}
\label{lemma:translate}
For $\mathcal{A}(\ket{\phi}, \ket{\psi})=\overline{\bra{\phi}} J_N \ket{\psi}$,
$\mathcal{C}_n(F_0 \ket{\phi},F_0 \ket{\psi})=
\mathcal{A}(\ket{\phi},\ket{\psi})$ for all $\ket{\phi},\ket{\psi} \in\
\mathcal{H}_n$.
\end{lemma}

\begin{proof}
$\mathcal{C}_n(F_0 \ket{\phi}, F_0 \ket{\psi}) =
\overline{\bra{\phi}} F_0^T (-i\sigma^y)^{\otimes n} F_0 \ket{\psi}$.
Now $F_0 J_N F_0^T = (-i \sigma^y)^{\otimes n}$
\cite[PropII.14]{BullockBrennen:03},
whence $F_0^T (-i \sigma^y)^{\otimes n} F_0 = J_N$.
\end{proof}

Now $Sp(N/2)$ is that copy of the symplectic group which embeds within
$SU(N)$ as the symmetries of $\mathcal{A}(-,-)$, i.e.
satisfying $\mathcal{A}(v\ket{\phi},v\ket{\psi})=\mathcal{A}(\ket{\phi},
\ket{\psi})$ for all $\ket{\phi},\ket{\psi} \in \mathcal{H}_n$.  In block form:
\begin{equation}
\label{eq:symplectic}
Sp(N/2) = \{ v \in SU(N) \; ; \; v^T J_N v = J_N \} =
\Bigg\{
\left(
\begin{array}{rr}
A & B \\
C & D \\
\end{array}
\right) \in SU(N)
\; ; \; 
\begin{array}{l}
A^T C \mbox{ is symmetric}, B^T D \mbox{ is symmetric}, \\
A^TD-C^TB=I
\end{array}
\Bigg\}
\end{equation}
As $E_0\; SO(N) \; E_0^\dagger = K_{2p}$, 
so too $F_0 \; Sp(N/2) \; F_0^T = K_{2p-1}$.

We next associate $\lambda_c(v)$ to $\mbox{spec}(a^2)$ for
$v=k_1 a k_2$ in the odd-qubit case.
Suppose we label $D$ to be the following diagonal subalgebra of $SU(N)$:
\begin{equation}
D=\Bigg\{\  
\sum_{j=0}^{N/2-1} d_j (\ket{j}\bra{j}+\ket{N/2+j}\bra{N/2+j}) \; ; \; 
\prod_{j=0}^{N/2-1} d_j = \pm 1 \ \Bigg\}
\end{equation}
Now there is a standard $SU(N)=KAK$ decomposition which follows
from $\theta_{\bf AII}(iH)=J_N(-iH^T)J_N^\dagger$ 
\cite[\S X.2,pg.452]{Helgason:01} and $\mathfrak{a}=\log D$
as above.  Given a $v \in SU(2^{2p-1})$, it writes 
$v=\omega_1 d \omega_2$, with $\omega_j \in Sp(N/2)$, $j=1,2$ and $d \in D$.

Suppose given $v \in SU(2^{2p-1})$, we then write
$F_0^T v F_0 = \omega_1 d \omega_2$, with $\omega_j \in Sp(N/2)$,
$j=1,2$ and $d \in D$.
The odd-qubit CCD again follows by a similarity transform:
$v=(F_0 \omega_1 F_0^T) (F_0 d F_0^T) (F_0 \omega_2 F_0^T)$ with
$a=F_0 d F_0^T \in A$, $k_j = F_0 \omega_j F_0^T \in K$, $j=1,2$ is a CCD.
Note that $a$ is diagonal on the GHZ-like basis
states $\{F_0 \ket{j}\}$.

\begin{lemma}
Let $n=2p-1$.   Then for $v=(F_0 \omega_1 F_0^T) 
(F_0 d F_0^T) (F_0 \omega_2 F_0^T)$ the CCD as above
with $d=\sum_{j=0}^{N/2-1} d_j (\ket{j}\bra{j}+\ket{N/2+j}\bra{N/2+j})$
diagonal and determinant one, we have
$\lambda_c(v)= \{ d_j^2 \}_{j=0}^{N/2-1} \sqcup \{ d_j^2 \}_{j=0}^{N/2-1}$
(counted with multiplicity.)
\end{lemma}

\begin{proof}
Given $A,B$, invertible, $\mbox{spec}(AB)=\mbox{spec}(BA)$.
Also, Equation \ref{eq:K_is_symmetry_group} is equivalent to
a matrix equation 
$k^T (-i \sigma^y)^{\otimes n} k= (-i \sigma^y)^{\otimes n}$ for all
$k \in K$.  Recall $F_0 J_N F_0^T=(-i \sigma^y)^{\otimes n}$.  Then
\begin{equation}
\begin{array}{l}
\lambda_c(v) \ = \ \mbox{spec}
\bigg( [(-i \sigma^y)^{\otimes n}]^\dagger v (-i \sigma^y)^{\otimes n} v^T
\bigg) 
\ = \ \mbox{spec}
\bigg( [(-i \sigma^y)^{\otimes n}]^\dagger k_1 a k_2
(-i \sigma^y)^{\otimes n} k_2^T a^T k_1^T
\bigg) \\
\  = \ \mbox{spec}
\bigg(  k_1^T [(-i \sigma^y)^{\otimes n}] k_1 a k_2
[(-i \sigma^y)^{\otimes n}]^T k_2^T a^T \bigg)
\  = \ \mbox{spec}
\bigg(  k_1^T [(-i \sigma^y)^{\otimes n}] k_1 a [k_2^T
(-i \sigma^y)^{\otimes n} k_2]^T a^T
\bigg) \\
\ =  \  \mbox{spec}
\bigg(  [(-i \sigma^y)^{\otimes n}] a [
(-i \sigma^y)^{\otimes n}]^T a^T
\bigg)
\  = \ \mbox{spec}
\bigg( - [F_0 J_N F_0^T] F_0 d F_0^T [F_0 J_N F_0^T] 
F_0 d^T F_0^T \bigg) \\
\ = \  \mbox{spec}(\; - F_0 J_N d J_N d^T F_0^T \; ) 
\ = \  \mbox{spec} (\; - J_N d J_N d \; ) 
\ = \ \mbox{spec} ( d^2) \\
\end{array}
\end{equation}
The last equality makes use of $d \in D$ repeat diagonal.
\end{proof}

\subsection{A convex hull argument in odd qubits}

\begin{definition}
\label{def:con_spec}
Suppose $n=2p-1$.  The \emph{reduced concurrence spectrum} 
$\tilde{\lambda}_c(v)$ 
of $v \in SU(N)$ is the set $\{ \lambda_j \}_{j=0}^{N/2-1}$
for $v=k_1 (F_0 d F_0^T) k_2$ a canonical decomposition of $v$
and $d=\sum_{j=0}^{N/2-1} \sqrt{\lambda_j} (\ket{j}\bra{j}
+\ket{N/2+j}\bra{N/2+j})$.
The \emph{convex hull} ${\tt CH}[\tilde{\lambda}_c(v)]$
of $\tilde{\lambda}_c(v)$ is the set 
convex linear combinations of the points of $\tilde{\lambda}_c(v)$, i.e.
\begin{equation}
{\tt CH}[\tilde{\lambda}_c(v)] = \Bigg\{ \ \sum_{j=0}^{N/2-1} 
t_j \lambda_j \ ; \ 0 \leq t_j \leq 1, \ \sum_{j=0}^{N/2-1} t_j =1 ,
\  \lambda_j \in \tilde{\lambda}_c(v) \ \Bigg\}
\end{equation}
\end{definition}

\vbox{
\begin{proposition}
Suppose $n=2p-1$ is an odd number of qubits.  Throughout, label 
$z_1 = \sum_{j=0}^{N-1} a_j \ket{j}$,
$z_2 = \sum_{j=0}^{N-1} b_j \ket{j}$, and
$\tilde{\lambda}_c(v)=\{ \lambda_j \}_{j=0}^{N/2-1}$.  Then the following hold.
\begin{itemize}
\item $\kappa_n(v)=\mbox{max } \bigg\{ \; \bigg| \sum_{j=0}^{N/2-1} \lambda_j 
(a_{N/2+j}b_j - a_j b_{N/2+j}) \bigg| \; ; \;
z_1^T J_N z_2 = 0, \ z_1^\dagger z_1 = z_2^\dagger z_2 =1 \; \bigg\}$
\item $(\kappa_n(v)=1) \Longleftrightarrow 
(\; 0 \in {\tt CH}[\tilde{\lambda}_c(v)] \;)$.
\end{itemize}
\end{proposition}
}

\begin{proof}
The first item follows from Lemma \ref{lemma:translate}, substituting
$x=F_0 z_1$, $y=F_0 z_2$.  We continue to the next item.

For the second item, we first prove $\Longrightarrow$.  If $\kappa_n(v)=1$,
then we may choose $z_1$, $z_2$ so that
\begin{equation}
\label{eq:astshriek}
\begin{array}{lclcl}
1 & = & |\sum_{j=0}^{N/2-1} \lambda_j (a_{N/2+j}b_j - a_j b_{N/2+j}) | 
& \leq & \sum_{j=0}^{N/2-1} |a_{N/2+j}b_j - a_j b_{N/2+j}| \\
& \leq & \sum_{j=0}^{N/2-1} \sqrt{|a_j|^2 + |a_{N/2+j}|^2} 
\sqrt{|b_j|^2 + |b_{N/2+j}|^2}
& \leq & 1 \\
\end{array}
\end{equation}
Here, note that the second inequality is an iterate
of 
\(
\mathcal{C}_1(\ket{\phi},\ket{\psi}) \leq \sqrt{\bra{\phi} \phi \rangle
\bra{\psi} \psi \rangle}, \mbox{ for all  }
\ket{\phi}, \ket{\psi} \in \mathcal{H}_1
\).  The last inequality in Equation \ref{eq:astshriek}
is the Schwarz inequality.

Now label $\alpha_j = a_{N/2+j}b_j - a_j b_{N/2+j}$, for
$0 \leq j \leq N/2-1$.  Then by Equation \ref{eq:astshriek},
\begin{equation}
\label{eq:is_one}
1 = | \sum_{j=0}^{N/2-1} {\lambda}_j \alpha_j| = 
\sum_{j=0}^{N/2 -1} |{\lambda}_j \alpha_j| = \sum_{j=0}^{N/2-1} |\alpha_j|
\end{equation}
Thus there must exist some $z \in \mathbb{C}$, $z \overline{z}=1$, so that
$\lambda_j \alpha_j = z |\alpha_j|$, and moreover
$\sum_{j=0}^{N/2-1} |\alpha_j|=1$.  On the other hand,
$z_1^T J_N z_2 =0$ demands that
$0 = \sum_{j=0}^{N/2-1} \alpha_j = 
z \sum_{j=0}^{N/2-1} |\alpha_j| \overline{\lambda}_j$.  
Multiplying by $\overline{z}$
and taking the complex conjugate,
\( 0 = \sum_{j=0}^{N/2-1} |\alpha_j| \lambda_j \)
which given $\sum_{j=0}^{N/2-1}|\alpha_j|=1$ by Equation
\ref{eq:is_one}
demands $0 \in {\tt CH}[\tilde{\lambda}_c(v)]$.

Consider now the converse case, i.e. $0 \in {\tt CH}[\tilde{\lambda}_c(v)]$.
Then there exist $t_j$ real, nonnegative so that
$0=\sum_{j=0}^{N/2-1} t_j \lambda_j$.  For $0 \leq j \leq N/2-1$, label
complex numbers $\alpha_j = t_j \overline{\lambda}_j$, so that
we have $1= \sum_{j=0}^{N/2-1} |\alpha_j|$ and moreover
$0=\overline{0}=\sum_{j=0}^{N/2-1} 
t_j \overline{\lambda}_j=\sum_{j=0}^{N/2-1} \alpha_j$.
We are reduced to the following question:  May we choose
$\{ a_j\}_{j=0}^{N-1}$, $\{ b_j \}_{j=0}^{N-1}$ so that
\begin{equation}
\alpha_j = a_{N/2+j}b_j-a_j b_{N/2+j} \mbox{\quad and \quad}
\sum_{j=0}^{N-1} |a_j|^2 = \sum_{j=0}^{N/2-1} |b_j|^2 = 1
\end{equation}

To do this, write $\alpha_j = |\alpha_j| \mbox{e}^{i \mbox{\footnotesize arg }
 \alpha_j}$,
and take $a_j = \sqrt{|\alpha_j|}$, $a_{N/2+j}=0$,
$b_j=0$, and $b_{N/2+j}=-\mbox{e}^{i \mbox{\footnotesize arg }\alpha_j} 
\sqrt{|\alpha_j|}$.
Then we see that $a_{N/2+j}b_j - a_j b_{N/2+j} = \alpha_j$.  Moreover,
\begin{equation}
|a_j|^2+|a_{N/2+j}|^2 = |\alpha_j|, \quad
|b_j|^2+|b_{N/2+j}|^2 = |\alpha_j| \quad
\mbox{ and }
\quad \sum_{j=0}^{N/2-1}|\alpha_j|=1
\end{equation}
Thus the vectors $z_1$, $z_2$ per the statement of the proposition are
normalized to be norm one.
\end{proof}

Hence, a similar picture emerges for concurrence capacity one for the
even and odd-qubit cases with these definitions.
The new feature, doubly-degenerate eigenvalues in $\lambda_c(v)$
arising from the $D$ above required
for type {\bf AII}, will \emph{a posteriori}
be an instance of Kramers' degeneracy; see \S \ref{sec:time}.

\begin{corollary}  
\label{cor:odd_prob_1}
For $n=2p-1$, \(
\mbox{lim}_{p \mapsto \infty} \ \ {\Large da}
(\; \{ \; a \in A \; ; \; \kappa_{n}(a)=1 \; \}) = 1 \).
\end{corollary}

The proof of the Corollary follows
by considering  probability density functions on the unit circle
\cite{BullockBrennen:03}, given that the number of concurrence
eigenvalues grows exponentially with $n$. 
Thus most unitary evolutions for large $n$ (of either
parity) are maximally entangling
as measured by concurrence.  It would be interesting 
but technically challenging to restate this
in terms of Haar measure $du$ on $SU(N)$.  The difficulty is
that the pullback measure from the $K \times A \times K$ to $SU(N)$ is
singular, namely singular near the set where the $A$ factor is an identity.
For future reference, we
summarize the concurrence capacity results of this section.

\vbox{
\begin{theorem}
\label{thm:kappas}
Let $\kappa_n(v)$, $\tilde{\kappa}_n(v)$ be the pairwise concurrence capacity
and concurrence capacity respectively.
\begin{enumerate}
\item  
\label{item:equal}
The pairwise capacity and the capacity are equal in any even
number of qubits.  Thus,
\begin{equation}
\tilde{\kappa}_n(v) = 
\left\{
\begin{array}{lr}
{\kappa}_n(v), & n=2p \mbox{ even} \\
0, & n=2p-1 \mbox{ odd} \\
\end{array}
\right.
\end{equation}
\item  
\label{item:symm}
For $n$ either even or odd, any CCD
by $v=k_1 a k_2$ satisfies $\kappa_n(v=k_1 a k_2)=\kappa_n(a)$.
\item  
\label{item:geq}
For any $n$, 
we must have $\kappa_{n+1}(v\otimes I_2) \geq \kappa_{n}(v)$.
\item  
\label{item:prob}
Suppose $n=2p-1$ is odd.  Then for $da$ the Haar measure on $A$,
\begin{equation}
\lim_{p \mapsto \infty}
\mbox{Prob}(\; \kappa_n(a)=1\; ) \ = \  
\lim_{p \mapsto \infty} da ( \{a \in A\; ; \; \kappa_n(a)=1\}) \ = \ 1
\end{equation}
\end{enumerate}
\end{theorem}
}


\section{An Algorithm Computing the Odd-qubit CCD}

In this section, we close a gap in the literature.  Specifically, we
present an algorithm for computing the CCD when the number of
qubits is odd.  
We make use of an algorithm \cite{DongarraEtAl:84}
by Dongarra, Gabriel, Koelling, and Wilkinson
cited in a survey \cite{chart:92} of diagonalization arguments.
The algorithm \cite{DongarraEtAl:84}, which appears in the
numerical matrix analysis literature,
improves the numerical stability and computational
efficiency of the earlier work on time reversal by Dyson \cite{Dyson:61}.

Recall from \S \ref{subsec:spectra_&_AII} that it suffices to compute the
standard type {\bf AII} $KAK$ decomposition given by
$SU(N)=Sp(N/2) \; D \; Sp(N/2)$ with $D$ the repeat diagonal subgroup
of $SU(N)$.  For given $v \in SU(2^{2p-1})$ for which we wish to
compute the CCD, suppose we obtain
$F_0^T v F_0 = \omega_1 d \omega_2$, with $\omega_j \in Sp(N/2)$,
$j=1,2$ and $d \in D$.  Then $v$ will have CCD
$v=k_1 a k_2 = (F_0 \omega_1 F_0^T) (F_0 d F_0^T) (F_0 \omega_2 F_0^T)$.
Before computing $SU(N)=Sp(N/2) \; D \; Sp(N/2)$, we make one new definition.

\begin{definition}
Let $H \in \mathbb{C}^{N \times N}$ be Hermitian.  Recall
$J_N=(-i \sigma^y) \otimes I_{N/2}$.  We say that the Hamiltonian
$H$ is \emph{$J_N$-skew symmetric} iff $HJ_N-J_NH^T=0$.
\end{definition}

\begin{remark}
In \cite{DongarraEtAl:84}, the above is the
definition of ``$H$ has a time reversal symmetry.''
Indeed, time reversal symmetry follows for the operator
$\Theta= J_N \tau$, ($\tau$ complex conjugation,) per the
upcoming Definition \ref{def:time_symmetry}.
Moreover, for the standard type {\bf AII} 
Cartan involution \cite[pg.452]{Helgason:01}
$\theta_{\bf AII}(X)=J_N \overline{X} J_N^T$, let
$\mathfrak{su}(N)=\mathfrak{p}_{\bf AII} \oplus \mathfrak{k}_{\bf AII}$
for the corresponding Cartan decomposition into $-1$ and $+1$ eigenspaces.  
Then $H$ is $J_N$ skew-symmetric iff. $iH \in \mathfrak{p}_{\bf AII}$.
Indeed on $\mathfrak{su}(N)$, $\overline{X}=-X^T$.  Hence
$-iH = J_N \overline{iH} J_N^T = - J_N iH^T J_N^T$ if and only
if $H J_N = J_N H^T$.
\end{remark}

\subsection
{Algorithm for the standard {\bf AII} $KAK$ decomposition,
$SU(N)=Sp(N/2) \; D \; Sp(N/2)$:}

The outline below for computing the standard $SU(N)=KAK$
decomposition of type {\bf AII} (see \S \ref{subsec:spectra_&_AII})
is similar to the {\bf AI} case used in \cite{BullockBrennen:03}
to compute the even-qubit CCD.  The added difficulties
are (i) a more complicated formula for $p^2$ and (ii) a more delicate
diagonalization argument for $p^2$ once computed.  In fact, the
latter requires the symplectic diagonalization argument referenced above.

\begin{lemma}
\label{lem:AIIpolar}
Suppose $v \in SU(N)$ with 
$v=pk$ for $p=\mbox{exp}(iH)$ with $H$ a $J_N$ skew-symmetric 
Hamiltonian and $k \in Sp(N/2)$.  Then $p^2 = -v J_N v^T J_N$.
\end{lemma}

\begin{proof}  We have
$H^T=-J_N H J_N$, given $J_N^\dagger=J_N^T = -J_N$.  Thus for any
$t \in \mathbb{R}$,
\(
[\mbox{exp}(iHt)]^T = J_N^\dagger \mbox{exp}(iHt) J_N = 
-J_N \mbox{exp}(iHt) J_N
\)
This holds in particular for $p$.
Now put $w=v^\dagger$, so that $w=\tilde{k} \tilde{p}$ for
$\tilde{k}=k^\dagger$, $\tilde{p}=p^\dagger$.
Thus
$\tilde{p}^T = J_N^\dagger \tilde{p} J_N$.  Moreover, 
$k \in Sp(N/2)$ demands
$\tilde{k}^T J_N \tilde{k} = J_N$, as $Sp(N/2)$ is a group.
Thus \( -J_N w^T J_N w = \tilde{p}^2\).  Taking the adjoint of each side
produces the result.
\end{proof}

With this lemma, we now present the algorithm for computing the standard
type {\bf AII} decomposition.
\begin{enumerate}
\item  Suppose $v=pk$ per Lemma \ref{lem:AIIpolar}.
Compute $p^2 = -v J_N v^T J_N$.
\item  We may write $p=\mbox{exp}(iH)$ for
some $J_N$ skew-symmetric
Hamiltonian $H$.
Compute a logarithm of
$p^2 = \mbox{exp}(2 i H)$.  The diagonalizing matrix implicit
in computing the matrix $\log$ need not be symplectic, and generic
logarithms will take the form $2iH$ for some $(2)H$ which is
$J_N$ skew-symmetric.
\item  
\label{step:dongarra}
Compute a symplectic matrix $\omega_1 \in Sp(N/2)$ so that
$iH_2=\omega_1^\dagger (iH) \omega_1$ is repeat diagonal,
per \S \ref{sec:sym_diag}.
\item  Label $p = \omega_1 \mbox{exp}(iH_2) \omega_1^\dagger$ and
$d=\mbox{exp}(iH_2)$.  Compute
$\omega_3=p^\dagger v$.  Then $\omega_3 \in Sp(N/2)$.
\item  Put $\omega_2 = \omega_1^\dagger \omega_3 \in Sp(N/2)$.  
Note that $\omega_1 d \omega_1^\dagger =p$.  Thus the 
type {\bf AII} decomposition is
\(
v = [\omega_1] [d] [\omega_1^\dagger \omega_3]=\omega_1 d \omega_2
\)
\end{enumerate}
This concludes the overview of computing 
\hbox{$SU(N)\; =\; Sp(N/2) \; D \; Sp(N/2)$}.  The
next subsection details Step \ref{step:dongarra}.

\subsection{Symplectic diagonalization}
\label{sec:sym_diag}

In this section we address the problem of finding the
eigendecomposition of a matrix $\mxH$ which is $J_N$ skew-symmetric.
Generically, these techniques work on any square matrix with an
even number of rows and columns, and there are no simplifications when
the size is a power of two.  Thus we describe the generic case where
$J_{2\ell} = \left( \begin{array}{rr} {\bf 0} & -I_\ell \\
I_\ell & {\bf 0} \\ \end{array} \right)$ and $H=H^\dagger$
is also $J_{2\ell}$ skew symmetric.

Explicitly, $J_{2 \ell}$-skew symmetric means
\(
 \mxH =  \bmx{c c}
          \mxA     &   \mxB    \\ 
      -\conj{\mxB} & \conj{\mxA} \emx \, ,
\)
where
$\mxA = \conjtransp{\mxA}$ and $\mxB = -\transp{\mxB}$
are $\ell \times \ell$ matrices.
We will construct a unitary skew-symmetric Hamiltonian
matrix $\mxW$ of the form
\(
\mxW = \bmx{c c}
         \mxU  & \mxV \\
      -\conj{\mxV} & \conj{\mxU} \emx \, 
\)
, so that the columns of $\mxW$ are the (right) eigenvectors
of $\mxH$.  Each eigenvalue $\lambda_k$ for $k=1,\dots,\ell$
of $\mxH$ is real and of multiplicity 2.  In particular, 
both the $k$th and the $(\ell+k)^{\mbox{th}}$ columns of $\mxW$ are 
eigenvectors of $\mxH$ corresponding to $\lambda_k$.
Also, given the block form, $\mxW \in Sp(N/2)$ up to global
phase.  

The algorithm of Dongarra et. al.
\cite{DongarraEtAl:84} proceeds in two major steps.  
First we reduce $\mxH$
to block diagonal form using a similarity transformation,
and then we use the QR algorithm to find the eigenvalues of the
blocks.  We consider each of these phases in turn.

First, we construct a skew-symmetric Hamiltonian
unitary matrix $\mxQ$ of the form
\(
\mxQ = \bmx{c c}
      \mxQ_1 & \mxQ_2 \\
      -\conj{\mxQ_2} & \conj{\mxQ_1} \\
\emx
\quad \mbox{ so that } \quad
\mxQ \mxH \conjtransp{\mxQ} =
\bmx{c c}
\mxT & 0 \\
   0 & \mxT \emx
\)
where $\mxT$ is real, symmetric, and tridiagonal.
We initialize $\mxQ$ to be the $2\ell \times 2\ell$
identity matrix.
In order to preserve the structure, we construct $\mxQ$
as the product of two simple types of matrices:
\begin{itemize}
\item
The product of $2 \times 2$ skew-symmetric Hamiltonian
matrices is also skew-symmetric Hamiltonian, and if
we let $r^2 = |a|^2 + |b|^2$, then a matrix
of the form
\(
\bmx{c c} \conj{a}/r & -b/r \\ \conj{b}/r & a/r \emx
\)
is unitary.
In addition, 
\(
\bmx{c c} \conj{a}/r & -b/r \\ \conj{b}/r & a/r \emx
\bmx{c c} a & b \\ - \conj{b} & \conj{a} \emx
=
\bmx{c c} r & 0 \\ 0 & r \emx
\)
so the unitary matrix can be used to introduce zeros.
Choose $j$ between $1$ and $\ell$ and
construct a matrix $\mxR$ as the $2\ell \times 2\ell$
identity matrix except that entries
$\mxR_{\ell+j,\ell+j} = \conj{\mxR_{j,j}} = a/r$
and
$\mxR_{j,\ell+j} = -\conj{\mxR_{\ell+j,\ell+j}} = -b/r$.
Then the product $\mxR \mxH$ is equal to $\mxH$ except
that the entries in rows $j$ and $\ell+j$ become
\begin{equation}
\bmx{c c} (\mxR \mxH)_{j,k} & (\mxR \mxH)_{j,\ell+k} \\
(\mxR \mxH)_{\ell+j,k} & (\mxR \mxH)_{\ell+j,\ell+k} \emx =
\bmx{c c} \conj{a}/r & -b/r \\ \conj{b}/r & a/r \emx
\bmx{c c} \mxA_{j,k} & \mxB_{j,k} \\ 
- \conj{\mxB_{j,k}} & \conj{\mxA_{j,k}} \emx \, ,
\end{equation}
$k=1,\dots,\ell$.  Since this product is skew-symmetric
Hamiltonian, so is $\mxR \mxH$, and it can be shown in
a similar way that $(\mxR \mxH) \conjtransp{\mxR}$
is skew-symmetric Hamiltonian.  Thus we can use $\mxR$
as a similarity transformation that preserves the
structure.
\item 
Let $\mxS$ be a real orthogonal matrix of dimension $\ell \times \ell$.
Then
\begin{equation}
\bmx{c c} \mxS & 0 \\ 0 & \mxS \emx
\bmx{c c} \mxA     &   \mxB    \\ -\conj{\mxB} & \conj{\mxA} \emx 
\bmx{c c} \conjtransp{\mxS} & 0 \\ 0 & \conjtransp{\mxS} \emx
=
\bmx{c c} \mxS \mxA \conjtransp{\mxS} &
\mxS \mxB \conjtransp{\mxS} \\
-\mxS \conj{\mxB} \conjtransp{\mxS} &
\mxS \conj{\mxA} \conjtransp{\mxS} \emx
\end{equation}
is skew-symmetric Hamiltonian.
\end{itemize}
Using these matrices, our construction takes $\ell-1$ steps.  We
describe the first step in detail.

The first step places zeros in the first column of the matrix in rows 
$3$ through $2\ell$.  To put a zero in position 
$(\ell+j,1)$ ($j=1,\dots,n$), we
construct an $\mxR$ matrix involving rows $j$ and $\ell+j$.
If $r_j^2 = |\mxA_{j,1}|^2 + |\mxB_{j,1}|^2$, then this 
matrix $\mxR_j$ is the identity matrix except that entries
$\mxR_{\ell+j,\ell+j} = \conj{\mxR_{j,j}} = \mxA_{j,1}/r_j$
and
$\mxR_{j,\ell+j} = -\conj{\mxR_{\ell+j,\ell+j}} = -\mxB_{j,1}/r_j$.
We replace $\mxH$ by $(\mxR \mxH) \conjtransp{\mxR}$
and update $\mxQ$ by premultiplying by $\mxR_j$, repeating
this for $j=1,\dots,\ell$.

We complete the first step by putting zeros in rows
$3$ through $\ell$ of column 1.  Note that these elements
are now real, since elements $2$ through $\ell$ are just
the values $r_j$. 
Thus we can construct a real orthogonal reflection (Householder)
matrix of the form $\mxS = \mxI - 2 s \transp{s}$ where
$\hat{s} = \transp{[0, r_2 + \|r\|, r_3, \dots, r_n]}$ and 
$s = \hat{s}/ \| \hat{s} \|$. A similarity transformation
of $\mxH$ by 
\( \bmx{c c} \mxS & 0 \\ 0 & \mxS \emx \)
produces the required zeros, and $\mxQ$ is updated by
premultiplying by this matrix.

Steps $2$ through $\ell-1$ are similar; in step $k$ we
first put zeros in the $\mxB$ portion of column $k$
using $\mxR$ matrices and then zero elements $k+2$
through $\ell$ of the $\mxA$ portion using a reflection
matrix.  The final result is that the transformed
$\mxH$ has a real tridiagonal matrix
$\mxT$ in place of $\mxA$ and $\conj{\mxA}$ and zeros
elsewhere.

The QR algorithm is considered to be the algorithm of choice for
determining all of the eigenvalues and eigenvectors of
a real symmetric tridiagonal matrix.  We use the algorithm to
form $X$, the matrix of eigenvectors of $T$.
Implementation of the
algorithm requires care, and high quality implementations
are available, for example, in LAPack \cite{lapack}.
Other codes are available at {\tt http://www.netlib.org}.

We construct the eigenvector matrices $\mxU$ and $\mxV$ as
$\mxU = \conjtransp{\mxQ_1} \mxVecs$
and $\mxV = \transp{\mxQ_2} \mxVecs$.
Note that most implementations of the QR algorithm do not
guarantee that the eigenvalues are ordered, so 
a final sort of the eigenvalues and the columns of $\mxU$
and $\mxV$ should be done at the end if desired.

\section{Time reversal, the CCD, and Kramers' nondegeneracy}
\label{sec:time}

The section presents three topics, all following from an
interpretation of $\mho$ from Equation \ref{eq:concurrence}
as a time reversal symmetry operator.  First, the Cartan
involution defining the CCD may be rewritten entirely in terms
of the spin-flip, and the eigenspaces of $\theta(iH)$ are associated
to time symmetric and antisymmetric Hamiltonians $H$ in a natural way.
Second, a well-known procedure exists to convert any $G=KAK$
decomposition into a polar decomposition, and the polar decomposition
associated to the CCD writes a unitary $v \in SU(N)$ as a product of
two factors, one evolution by a time symmetric Hamiltonian and one
evolution by a time anti-symmetric Hamiltonian.  Third, we demonstrate
the entangled eigenstates of Kramers' nondegeneracy as described in
the introduction and consider the perturbative stability of this
entanglement under time reversal symmetry breaking.

\subsection{Spin-flips as time reversal symmetry operators}

Recall the \emph{Bloch sphere} (e.g. \cite{NielsenChuang:00}),
which provides a picture of the data space of one qubit.
As a remark, the Bloch sphere may be thought of as
a parameterization of the complex projective line $\mathbb{CP}^1$
(e.g. \cite[\S40]{Munkres:84}.)
Briefly, $\mathbb{CP}^1$ is the set of all equivalence classes of
vectors in $\mathbb{C}^2$ up to multiple by a nonzero complex
scalar.  To associate such
a class with a Bloch vector, normalize $\ket{\psi}$ as above so as to write
$\ket{\psi}=r \mbox{e}^{it}[\cos \frac{\theta}{2} \ket{0}+
\mbox{e}^{i \varphi} \sin \frac{\theta}{2} \ket{1}]$.
The Bloch sphere vector of $\ket{\psi}$, say 
$[\ket{\psi}]$, is
given in spherical coordinates by $(1,\theta,\varphi)$ 
\cite[pg.15]{NielsenChuang:00}.
Recall also that the north pole is $[\ket{0}]$ 
and $[\ket{1}]$ is the south pole.

Now let $\vec{b} \in (\mathbb{F}_2)^n$ be an $n$-bit string.  The typical
procedure when quantizing a classical computation is to extend the classical
outputs linearly without phases.  Thus, a reasonable interpretation
of quantum bit-flip would be $(\sigma^x)^{\otimes n}$.  This is the common
interpretation, but note that in one qubit
$\sigma^x$ is not reflection on the Bloch sphere and
indeed has a fixed state, $(1/\sqrt{2})(\ket{0} + \ket{1})$.  
Rather, the odd reflection of a single qubit under the Bloch parameterization 
of $\mathbb{CP}^1$ is the spin-flip $\ket{\psi} 
\mapsto \overline{ (-i \sigma^y) \ket{\psi}}=(-i\sigma^y) 
\overline{\ket{\psi}}$.

The appropriate physical interpretation of the spin-flip 
is as a time reversal symmetry operator
\cite[Ch.26]{Wigner:59} \cite[pp.314-322]{Gottfried:66}
\cite{Kramers:64,Sakurai:85}.
Wigner defined a generic time reversal symmetry operator $\Theta$
as any $\mathbb{R}$-linear
involutive map of the quantum Hilbert space which is antiunitary,
i.e. complex anti-linear ($\Theta( \alpha \ket{\psi_1}+\beta \ket{\psi_2})=
\overline{\alpha} \Theta{\ket{\psi_1}} + \overline{\beta} 
\Theta{\ket{\psi_2}}$)
and orthogonal in the induced real inner-product on $\mathbb{R}^{2p}
\cong \mathbb{C}^p$.  
Generic time reversal symmetry operators
are usually denoted by a capital $\Theta$; we ask the reader's forebearance
in distinguishing this from the lower-case $\theta$ describing a Cartan
involution.

Such a time reversal symmetry operator $\Theta$
maps the state of a system to its 
motion-reversed state, so that momentum eigenstates 
transform as $\Theta\ket{\mathbf{p}}=\ket{\mathbf{-p}}$.
In particular, if our qubit is a spin $\frac{1}{2}$ particle,
e.g. with $\ket{0}=\ket{\uparrow}$ and $\ket{1}=\ket{\downarrow}$, 
then $\ket{\psi} \in \mathcal{H}_1$ rotates counterclockwise about
the Bloch sphere vector of $[\ket{\psi}]$.  Thus $\mho$
per Equation \ref{eq:concurrence} is the natural quantum angular
momentum reversal in $n$-qubits.  Indeed, the total 
spin angular 
momenum, $\vec{S}=\sum_{j=1}^n\vec{\sigma_j}$, is inverted under time
reversal:  $\mho \vec{S} \mho^{-1}=-\vec{S}$.
Spin-flip operators may be defined for $d$-level systems (qudits)
but may not both preserve pure states and commute with
local unitaries \cite{Rungta:01}.

We note in passing that the spin-flip picture also allows one
to quickly rederive one of the monotone properties.
Namely, antipodal points in the Bloch sphere
parameterization of the complex projective line $\mathbb{CP}^1$
correspond to Hermitian-orthogonal states of $\mathcal{H}_1$.
Hence, $C_n(\ket{\psi})=| \bra{\psi} \mho \ket{\psi}|=0$ if
$\ket{\psi}=\otimes_{j=1}^n \ket{\psi_j}$ (the monotone
property,) since in this event $\bra{\psi} \mho \ket{\psi}$ has a 
factor $\bra{\psi_j} \mho \ket{\psi_j}=0$.
More generally $C_n(\ket{\psi})=0$
whenever $\ket{\psi} = \ket{\psi_1} \otimes \ket{\psi_2}$ for
$\ket{\psi_1} \in \mathcal{H}_{n-1}$ and $[\ket{\psi_2}]$ a point
on the Bloch sphere.  However, the latter is not an equivalence
for $n$ even.  Consider 
$\ket{\mbox{W}_4} = (1/2) \big( \ket{0001}+\ket{0010}+\ket{0100}+
\ket{1000} \big)$.

\subsection{Time reversal and the CCD Cartan Involution}

We next show that physically, the eigenspaces of the Cartan
involution producing the CCD correspond to
$\mho$-time symmetric and $\mho$-time antisymmetric Hamiltonians.
They are then explicitly described in the Pauli-tensor basis
of $\mathfrak{su}(N)$ in much more compact form than in
Dirac notation \cite{BullockBrennen:03}.

\begin{definition}
\label{def:time_symmetry}
Consider $H$ a Hamiltonian on a finite dimensional Hilbert space
$\mathcal{H}$; i.e. $H$ is selfadjoint within 
$\mbox{End}_{\mathbb{C}}(\mathcal{H}) 
\subset \mbox{End}_{\mathbb{R}}(\mathcal{H})$.
Then $H$ is time reversal symmetric with respect to
$\Theta$ iff $H=\Theta H \Theta^{-1}$ as 
elements of $\mbox{End}_{\mathbb{R}}(\mathcal{H})$.  A Hamiltonian
is time reversal anti-symmetric with respect to
$\Theta$ iff $H=-\Theta H \Theta^{-1}$.
\end{definition}

\vbox{
\begin{proposition}
\label{prop:CCDisTR}
Let $\theta(X)$ per Definition \ref{def:CCD}.  Label 
$\mathfrak{su}(N)=\mathfrak{p} \oplus \mathfrak{k}$ as the
$-1$ and $+1$-eigenspaces of $\theta$.  Let $\mho$ be the spin-flip.
Then (i) for $H$ a traceless Hamiltonian, so that $iH \in \mathfrak{su}(N)$,
$\theta(iH)=\mho \; (iH) \; \mho^{-1}$, with the right-hand side viewed
as a composition of $\mathbb{R}$-linear maps.
Also (ii) $(H \mbox{ has time reversal symmetry with respect to } \mho)$
$\Longleftrightarrow$ $(iH \in \mathfrak{p},)$
and (iii) $(H \mbox{ has time reversal anti-symmetry with respect to } \mho)$
$\Longleftrightarrow$ $(iH \in \mathfrak{k})$.
\end{proposition}
}

\begin{proof}
Let $\tau$ denote the complex conjugation operator
$\ket{\psi} \mapsto \overline{\ket{\psi}}$.  Then
$\mho = (-i \sigma^y)^{\otimes n} \tau =
\tau (-i \sigma^y)^{\otimes n}$, given $-i \sigma^y$ real.
So $\mho^{-1}= \tau [(-i \sigma^y)^{\otimes n}]^\dagger$.  Moreover,
$[(-i \sigma^y)^{\otimes n}]^\dagger=(-I_N)^n (-i \sigma^y)^{\otimes n}$.  
Finally, $\tau (iH) \tau = \overline{iH}$.  Thus,
\begin{equation}
\mho \; (iH) \; \mho^{-1} \ = \ 
(-i \sigma^y)^{\otimes n} \tau \; (iH) \;
\tau [(-i \sigma^y)^{\otimes n}]^\dagger \ = \ 
(-I_N)^n (-i \sigma^y)^{\otimes n} \overline{(iH)}
[(-i \sigma^y)^{\otimes n} = \theta(iH)
\end{equation}
The latter two items follow at once.
\end{proof}

With the above proposition, we may describe the infinitesimal Cartan
decomposition $\mathfrak{su}(n)=\mathfrak{p} \oplus \mathfrak{k}$
directly in terms of tensors of Pauli operators.
Let $j$ denote either $0$, $x$, $y$, or $z$, with
$\sigma^j=I_2$ in case $j=0$ and Pauli matrices $\sigma^x$, $\sigma^y$,
or $\sigma^z$ as appropriate otherwise.  
A multiindex $J=j_1j_2 \cdots j_k \cdots j_n$
denotes a string of $n$, 
and $J$ will be said to be nonzero if some
$j_k \neq 0$.  Finally, let
$i \sigma^{\otimes J}$ denote $i \otimes_{k=1}^n (\sigma^{j_k})$.  Then
\(
\mathfrak{su}(N)= \oplus_{\mbox{\footnotesize all nonzero } J} 
\mathbb{R} \{ i \sigma^{\otimes J}\}
\)
We have the following corollary, discovered independently by
Bremner et. al. \cite[Thm.5]{BremnerEtAl:03}.

\vbox{
\begin{corollary}
\label{cor:Pauli}
Continue the convention of the previous paragraph, and write
\begin{equation}
\mathfrak{su}(N)= \bigg( \bigoplus_{\# J = 0 \; 
\mbox{\footnotesize mod } 2}
\mathbb{R} \{ i \sigma^{\otimes J}\} \bigg)
\ {\bigoplus} \ 
\bigg( \bigoplus_{\# J = 1 \; \mbox{\footnotesize mod } 2}
\mathbb{R} \{ i \sigma^{\otimes J}\} \bigg)
\end{equation}
The above is the infinitesimal Cartan decomposition of $\theta(iH)$, i.e.
\(
\mathfrak{p} \ = \  \bigoplus_{\# J = 0 \; \mbox{\small mod} 2}
\mathbb{R} \{ i \sigma^{\otimes J}\}
, \quad \mbox{ and } \quad
\mathfrak{k} \ = \  \bigoplus_{\# J = 1 \; \mbox{\small mod} 2}
\mathbb{R} \{ i \sigma^{\otimes J}\}
\)
In particular, $K$ is the Lie group of those unitaries which are exponentials
of Hamiltonians with time reversal anti-symmetry w.r.t. $\mho$.
\end{corollary}
}

\begin{proof}  Distinct Pauli matrices anti-commute, each has
$(\sigma^j)^2=I_2$, and
$\sigma^y$ is purely imaginary while $\sigma^x$, $\sigma^z$,
and $I_2$ are real.  Considering the tensors case by case 
completes the proof.
\end{proof}

\subsection{A time reversal polar decomposition}

We next consider the polar decomposition which may be derived from
the CCD.  In most treatments,
the polar decomposition of a general Cartan
involution is proven and then a $G=KAK$ theorem is derived from it.
We next use the CCD to produce a polar decomposition
for time reversal symmetry.  This practical decision avoids
rearguing the $G=KAK$ theorem for compact groups 
\cite[thm8.6,\S VII.8]{Helgason:01}.

\begin{corollary}  
\label{cor:antisymsym}
Suppose $v \in SU(N)$ is a phase normalized quantum
computation in $n$ qubits.  Then we may write
$v= \mbox{exp}(i H_{\mathfrak{p}}) \mbox{exp}(i H_{\mathfrak{k}})$ for
some Hamiltonians $H_{\mathfrak{p}}$, $H_{\mathfrak{k}}$ such that
$H_{\mathfrak{p}}$ has time reversal symmetry and
$H_{\mathfrak{k}}$ has time reversal anti-symmetry with respect to the
spin-flip $\mho$.
\end{corollary}

\begin{proof}
Let $v=k_1 a k_2$ be the CCD of $v \in SU(N)$.  Then in particular
$v=(k_1 a k_1^\dagger) (k_1 k_2)$.
Since $K$ is a group, $k_1 k_2$ is a time
antisymmetric evolution by Proposition
\ref{prop:CCDisTR}.  Moreover, let $a = \mbox{exp} \; iH$
for $iH \in \mathfrak{a} \subset \mathfrak{p}$ a time symmetric Hamiltonian.
As $iH \in \mathfrak{p}$, we have
$\theta(iH)=[(-i \sigma^y)^{\otimes n}]^\dagger \overline{(iH)}
(-i\sigma^y)^{\otimes n}=-iH$.  Moreover, 
$k \in K$ is a symmetry of the concurrence form 
(Eq. \ref{eq:K_is_symmetry_group},)
which as a matrix equation demands $k^T(-i \sigma^y)^{\otimes n} k=
(-i \sigma^y)^{\otimes n}$.  Hence
$k_1^T (-i \sigma^y)^{\otimes n} = (-i\sigma^y)^{\otimes n}k_1^\dagger$, and
for $k_1 iH k_1^\dagger \in \mathfrak{p}$:
\begin{equation}
\theta(k_1 iH k_1^\dagger)=[(-i \sigma^y)^{\otimes n}]^\dagger \overline{k}_1
\overline{(iH)} k_1^T (-i \sigma^y)^{\otimes n} =
k_1 [(-i \sigma^y)^{\otimes n}]^\dagger \overline{(iH)} 
(-i \sigma^y)^{\otimes n}
k_1^\dagger = -k_1 (iH) k_1^\dagger
\end{equation}
Thus $k_1 (iH) k_1^\dagger$ has time reversal symmetry, and the
usual matrix exponential formula (valid since $SU(N)$ is linear) shows
$k_1 a k_1^\dagger=\mbox{exp}[k_1 (iH) k_1^\dagger]$.
\end{proof}

\begin{remark}  Note that the vector space decomposition 
$\mathfrak{su}(N)=\mathfrak{p}\oplus \mathfrak{k}$
makes clear any such $v$ may be approximated by rapid pulsing of the time
symmetric and anti-symmetric factors, by applying the
Trotter formula (e.g. \cite[\S4.7.2]{NielsenChuang:00}.)  However,
the decomposition above requires no such pulsing of the
time-symmetric and time-antisymmetric Hamiltonians.
\end{remark}

\subsection{Kramers' nondegeneracy}

Finally, we rederive Kramers' degeneracy in the case of $\mho$
and note a further, $\mho$-specific nondegeneracy property.
Recall Kramers' 
degeneracy \cite{Kramers:30,Kramers:64} 
proves that the eigenstates of a collection of 
an odd number of spin $\frac{1}{2}$ electrons become doubly degenerate 
in the exclusive presence of a time-symmetric interaction,
such as an electric field.  The degeneracy is broken 
with the introduction of a magnetic field.  In terms of an energy Hamiltonian
$H$ of the system, the degeneracy corresponds to $2$ or greater dimensional
eigenspace for energy eigenstates.

\begin{lemma}
\label{lem:flipground}
Suppose that $\ket{\psi} \in \mathcal{H}_n$ is an eigenstate of some
traceless Hamiltonian $H$ which has time reversal symmetry, with
eigenvalue $\lambda \in \mathbb{R}$.  Then the spin-flip
$\mho \ket{\psi}$ is also an eigenstate of eigenvalue $\lambda$.
\end{lemma}

\begin{proof}
Since $iH$ has time reversal symmetry, $\theta(iH)=-iH$.  Thus
$(-i \sigma^y)^{\otimes n} (iH) + \overline{(iH)} (-i \sigma^y)^{\otimes n}
=0$, and taking a complex conjugate produces
\(
(-i \sigma^y)^{\otimes n} \overline{(iH)} + 
{(iH)} (-i \sigma^y)^{\otimes n}=0
\).
Now $(iH) \ket{\psi}=\lambda \ket{\psi}$, so that
\begin{equation}
(iH) \mho \ket{\psi} = 
(iH) (-i\sigma^y)^{\otimes n} \overline{\ket{\psi}} =
-(-i \sigma^y)^{\otimes n} \overline{(iH) \ket{\psi}} =
-(-i \sigma^y)^{\otimes n} \overline{ i \lambda \ket{\psi}} =
i \lambda \mho \ket{\psi}
\end{equation}
This concludes the proof.
\end{proof}

\begin{theorem}
\label{thm:Kramers}
{\bf (Cf. Kramers' degeneracy, 
\cite{Kramers:30,Kramers:64}\cite[pg.281]{Sakurai:85})}
Let $H$ be a traceless Hamiltonian on some number $n$ of quantum-bits.
Suppose $H$ has time reversal symmetry with respect to $\mho$.
Let $\lambda$ be a fixed eigenvalue of $H$.
Then either (i) $\lambda$ is degenerate
with even multiplicity or 
(ii) the normalized eigenstate $\ket{\lambda}$
has $C_n(\ket{\lambda})=1$.  For $n$ odd, case (i) holds:
all $\lambda$ are degenerate with even multiplicity.
\end{theorem}

\begin{proof}
Let $\lambda_j$ be some eigenvalue of $H$.
By Lemma \ref{lem:flipground}, both $\ket{\lambda_j}$ and
$\mho \ket{\lambda_j}$ are energy eigenstates.  Should these two states
be linearly independent, then $\lambda_j$ is degenerate.  If any eigenvalue is 
non-degenerate, say $\lambda_k$, then by
antiunitarity of $\mho$, we must have
$\mho \ket{\lambda_k} = \mbox{e}^{i \varphi} \ket{\lambda_k}$ for some
global phase $\varphi$.  Using $C_n(\ket{\lambda_k})=
|\bra{\lambda_k} \mho \ket{\lambda_k}|$ we see that 
this eigenstate must have concurrence one.

Suppose in particular $n=2p-1$.  Then $\mathcal{C}_n(-,-)$ is
antisymmetric and vanishes on the
diagonal, implying $\bra{\lambda_j} \mho \ket{\lambda_j}=0$
for all $j$.
Consequently, $\ket{\lambda_j}$ and $\mho \ket{\lambda_j}$ are Hermitian
orthogonal and may not be dependent, implying case (i).
\end{proof}

Thus, for the spin-flip $\mho$ there
is in addition to the Kramers' degeneracy a Kramers' nondegeneracy.
As always, if $n$ is odd so that the total $n$-qubit system is a fermion,
then a time reversal symmetric Hamiltonian implies that all energy
eigenstates are degenerate.  
Yet moreover in the specific case
of $\mho$ and \emph{$n$ even,} a nondegenerate eigenstate must also have
maximal concurrence and hence be entangled.

We provide some illustrative examples.  First note that there are 
many systems endowed with time reversal symmetric Hamiltonians. In 
particular, any system with (exclusively) 
pairwise nearest neighbor coupling between 
qubits has $iH\in\mathfrak{p}$, by Corollary
\ref{cor:Pauli}.  An example of an interaction that occurs 
in many solid state systems is the quantum XYZ model:
\begin{equation}
\label{eq:XYZ}
H_{XYZ}=\sum_{<j,k>}J_x\sigma^x_j\sigma^x_k + J_y\sigma^y_j\sigma^y_k
+ J_z\sigma^z_j\sigma^z_k
\end{equation}
with $J_{x,y,z}\in\mathbb{R}$ where 
the sum is taken over all nearest neighbor pairs and the boundaries 
may be fixed or periodic.  In one dimension, these nearest
neighbor coupled systems are known as spin chains.  Spin chain Hamiltonians
are of great theoretical interest, for under the appropriate parameter
regime they exhibit 
long range classical correlations near a quantum 
phase transition \cite{Lieb:61}.  
We can characterize the dynamics of entanglement in spin chains using the 
concurrence capacity.  With this goal in mind we observe the following 
useful fact:

\begin{proposition}
Let $\mathfrak{p}$, $\mathfrak{k}$ be as in Corollary \ref{cor:Pauli}.
If $iH \in \mathfrak{p}$ and $H \in \mathbb{R}^{N \times N}$, then 
$\lambda_c(u=e^{-iH t})=\{e^{-2i\lambda_j t}\}$ where 
$t\in\mathbb{R}$ parametrizes time and ${\lambda_j}\in\mathbb{R}$ are 
the eigenvalues of $H$.
\end{proposition}

\begin{proof}
By Definition \ref{def:conspec} the concurrence 
spectrum of the unitary generated 
by $iH$, $u=e^{-iHt}$ is 
\begin{equation}
\label{eq:realhamil}
\begin{array}{lclcl}
\lambda_c(u)& = & \mbox{spec}[\; (-i\sigma^y)^{\otimes n\dagger}e^{-iHt}
(-i\sigma^y)^{\otimes n}(e^{-iHt})^T \; ] 
& = &  \mbox{spec}(e^{-iHt}e^{-iH^Tt}) \\
& = &  \mbox{spec}(e^{-2iHt}) 
& = & \{\; e^{-2i\lambda_j t} \; ; \; \lambda_j \in \mbox{spec}(H) \; \} \\
\end{array}
\end{equation}  
We have used $(-i\sigma^y)^{\otimes n\dagger}
\overline{iH}(-i\sigma^y)^{\otimes n}=-iH$ and 
therefore $(-i\sigma^y)^{\otimes n\dagger}H(-i\sigma^y)^{\otimes n}=H$ 
because $H$ is real.  The third line is a consequence of $H$ being Hermitian.
\end{proof}

The quantum XYZ Hamiltonian has
time reversal symmetry with respect to the spin-flip
$\mho$.  We next demonstrate how to 
build up entanglement with such a system.  Consider a collection of $n$
qubits laid out in a cyclic array interacting under the 
Ising class of Hamiltonians given by $H_{XYZ}$ with $J_x=J_y=0$:
$H_{\mbox{\small Is}} =\sum_{j=1}^{n} J_z\sigma^z_j\sigma^z_{j+1}$,
where we identify $\sigma^z_{n+1}=\sigma^z_1$.

The eigenvalues are given by 
$\{\lambda_j\}=\{J_z(n-2\sum_kj_k\oplus j_{k+1});\ j=j_1j_2\ldots j_n\}$
\cite{Lieb:61}, where the addition 
is done modulo $2$ over the components $j_k$ of the binary expansion of $j$.  
For $n$ even, each eigenvalue $\lambda_j$ is paired with another of 
opposite sign and 
in particular, $\lambda_0=-\lambda_{N-1}$ with 
$|\lambda_0|=n|J_z|=\lambda^{\mbox{\footnotesize max}}$.  The concurrence 
spectrum of $u=e^{-iH_{Is}t}$ is composed of complex conjugate pairs 
and the concurrence capacity $\tilde{\kappa}_n(u)$ may be computed 
explicitly.  Then 
$\tilde{\kappa}_n(u)=
\mbox{max } \{ |\sum_{j=0}^{N-1} a_j^2 e^{-2i\lambda_j t}|\; ; \; 
z^\dagger z = 1, z^T z =0 \}$ where $z=\sum_{j=0}^{N-1}a_j\ket{j}$,
per Equation \ref{eq:max_kappa}.
Maximum capacity is obtained when the 
convex hull condition is satisfied which occurs 
precisely when the concurrence spectrum extends outside 
the right half of the complex plane.  The minimum time at which this 
occurs is given by $e^{-2i\lambda^{{max}}t_{{min}}}=i$ or 
$t_{{min}}=\pi/4|\lambda_0|=\pi/4n|J_z|$.  

The existence of a time reversal symmetry in the interaction between qubits 
gives us 
important information about the nature of quantum correlations in the 
energy eigenstates.  Applying Theorem \ref{thm:Kramers}, we immediately 
find that the ground state of a Hamiltonian $H$ with time reversal 
symmetry has maximum $n$-concurrence if it is unique.  Examples of 
interactions satisfying these conditions are the XYZ Hamiltonian with 
$(J_x=J_y=J_z=J>0)$, denoted the XXX Hamiltonian, and the XY Hamiltonian
$(J_x=J_y,J_z=0)$
\cite{Lieb:61}.
In particular, the XXX Hamiltonian with $J>0$ has been shown 
to have non-degenerate ground states in any number of dimensions, with 
or without periodic boundary conditions,
provided the underlying lattice has a reflection symmetry about some plane
(ibid.)  

To illustrate this phenomenon we consider what happens when the time reversal
symmetry is broken by adding a time-antisymmetric 
term to the  XY Hamiltonian:
\begin{equation}
H=\sum_{j=1}^{n}J(\frac{1+g}{4}\sigma^x_j\sigma^x_{j+1} + 
\frac{1-g}{4}\sigma^y_j\sigma^y_{j+1}) + \frac{h_z}{2}\sigma^z_j,
\end{equation}
where $\sigma^{\alpha}_{n+1}\equiv\sigma^{\alpha}_1$. The presence of the 
linear term proportional to the total spin projection 
operator $S_z=\sum_{j=1}^n \sigma^z_j$, breaks the time reversal symmetry 
so that $iH\notin\mathfrak{p}$ when $h_z\neq 0$.
For zero 
magnetic field and $0\leq g<1$, the 
Hamiltonian is time reversal symmetric and the ground state is non-degenerate 
meaning the concurrence is maximal.  In the isotropic case $(g=0)$, 
the Hamiltonian
commutes with $S_z$ and eigenstates are independent 
of $h_z$.  For magnetic field strengths below some critical value, 
$|h|<h_{crit}$ the ground 
state corresponds to an eigenstate with eigenvalue $s_z=0$ of the 
operator $S_z$.  This ground state has maximal concurrence.  For 
$|h_z|>h_{crit}$, the ground state corresponds to
an eigenvalue $s_z\neq 0$ and the concurrence is zero 
\cite{BrennenBullock:04}. 

\section{Conclusions}

We show that the odd-qubit concurrence canonical decomposition admits
generalizations of all constructions studied on the even qubit CCD.  
In particular, a generalized pairwise concurrence capacity
may be defined, and the operators for which this is maximal are characterized
by a convex hull condition on the concurrence spectrum.  Again for an
odd number of qubits, we find that for large odd $n$ most
unitaries have maximal concurrence capacities.
Moreover, we provide an explicit algorithm for computing the odd-qubit 
CCD.

These advances are complemented by new interpretation of the original
inputs to the $G=KAK$ theorem which define the CCD.
Specifically, they may be rewritten
in terms of time reversal symmetry $\mho$ which is
the spin-flip in $n$ quantum bits, and the CCD is best understood
in terms of such symmetries.  For example,
the odd-qubit CCD is a type {\bf AII} $KAK$ decomposition, and as such
much have degenerate eigenvalues.  In fact, this recaptures Kramers'
degeneracy for the odd-qubit spin-flip, and a more careful study of
the arguments reveals a Kramers' nondegeneracy:  Nondegenerate
eigenstates of $\mho$ time reversal symmetric Hamiltonians
only exist when the number of quantum bits is even and
\emph{moreover} must be highly entangled.
Specifically, such $\ket{\lambda}$ are
highly entangled in the sense that the concurrence
$C_n(\ket{\lambda}) = 
\big| \langle \lambda | \mho | \lambda \rangle \big| =1$.
Finally, the polar decomposition extracted from the
CCD in the usual way accomplishes the following:
any unitary $n$-qubit evolution is 
a product of precisely one time reversal symmetric and one
time reversal antisymmetric evolution.

\noindent
{\bf Acknowledgements}
SSB acknowledges a National Research Council postdoctoral fellowship,
GKB acknowledges support through DARPA QuIST, and DPO
acknowledges NSF Grant CCR-0204084.

\noindent
{\bf NIST disclaimer. }
  Certain commercial equipment or instruments may be identified in this
  paper to specify experimental procedures.
  Such identification is not intended to imply recommendation
  or endorsement by the National Institute of Standards and
  Technology.

\end{document}